\documentclass[useAMS, usegraphicx, usenatbib]{mn2e}
\usepackage{aas_macros}
\usepackage{amsmath}
\usepackage{graphicx}
\usepackage{natbib}

\newcommand{\runit}{\,h^{-1}\,\mathrm{Mpc}}
\newcommand{\runitk}{\,h^{-1}\,\mathrm{kpc}}
\newcommand{\munit}{\,h^{-1}\,\mathrm{M}_{\sun}}

\setcitestyle{authoryear,round,comma,aysep={},yysep={,},notesep={}}
\title[Baryons and the two-point correlation function]{The impact of baryonic processes on the two-point correlation functions of galaxies, subhaloes and matter}
\author[M. P. van Daalen et al.]{Marcel P. van Daalen$^{1,2}$\thanks{E-mail: daalen@strw.leidenuniv.nl}, Joop Schaye$^{1}$, Ian G. McCarthy$^{3}$, C. M. Booth$^{4}$
\newauthor and Claudio Dalla Vecchia$^{5}$\\\\
$^1$Leiden Observatory, Leiden University, P.O. Box 9513, 2300 RA Leiden, The Netherlands\\
$^2$Max Planck Institute for Astrophysics, Karl-Schwarzschild Stra\ss{}e 1, 85741 Garching, Germany\\
$^3$Astrophysics Research Institute, Liverpool John Moores University, 146 Brownlow Hill, Liverpool L3 5RF, United Kingdom\\
$^4$Department of Astronomy \& Astrophysics, The University of Chicago, Chicago, IL 60637, USA\\
$^5$Max Planck Institute for Extraterrestrial Physics, Gie\ss{}enbachstra\ss{}e 1, 85748 Garching, Germany}
\begin{document}
\pagerange{\pageref{firstpage}--\pageref{lastpage}} \pubyear{2012}
\maketitle
\label{firstpage}
\begin{abstract}
The observed clustering of galaxies and the cross-correlation of galaxies and mass provide important constraints on both cosmology and models of galaxy formation. Even though the dissipation and feedback processes associated with galaxy formation are thought to affect the distribution of matter, essentially all models used to predict clustering data are based on collisionless simulations. Here, we use large hydrodynamical simulations to investigate how galaxy formation affects the autocorrelation functions of galaxies and subhaloes, as well as their cross-correlation with matter. We show that the changes due to the inclusion of baryons are not limited to small scales and are even present in samples selected by subhalo mass. Samples selected by subhalo mass cluster $\sim 10\%$ more strongly in a baryonic run on scales $r \ga 1\runit$, and this difference increases for smaller separations. While the inclusion of baryons boosts the clustering at fixed subhalo mass on all scales, the sign of the effect on the cross-correlation of subhaloes with matter can vary with radius. We show that the large-scale effects are due to the change in subhalo mass caused by the strong feedback associated with galaxy formation and may therefore not affect samples selected by number density. However, on scales $r \la r_\mathrm{vir}$ significant differences remain after accounting for the change in subhalo mass. We conclude that predictions for galaxy-galaxy and galaxy-mass clustering from models based on collisionless simulations will have errors greater than $10\%$ on sub-Mpc scales, unless the simulation results are modified to correctly account for the effects of baryons on the distributions of mass and satellites.
\end{abstract}

\section{Introduction}
Many cosmological probes are used in order to derive the values of the parameters describing our Universe, often relying on some aspect of large-scale structure. By combining different probes, degeneracies can be broken and the constraints on the numbers that characterise our Universe can be improved. However, observations alone are not enough: strong theoretical backing is needed to interpret the data and to avoid, or at least to reduce, unexpected biases.

Modelling our Universe as a dark matter only $\Lambda$CDM universe was a reasonable approximation for the interpretation of past data sets. However, over the last few years it has become clear that for many probes this is no longer the case in the era of precision cosmology: ignoring processes associated with baryons and galaxy formation may lead to serious biases when interpreting data. The existence of baryons and the many physical processes associated with them have been shown to significantly impact, for example, the mass profiles \citep[e.g.][]{Gnedin2004,Duffy2010,Abadi2010,Governato2012,Martizzi2012,Velliscig2014} and shapes of haloes \citep[e.g.][]{Kazantzidis2004,Tissera2010,Bryan2013}, the clustering of matter \citep[e.g.][]{White2004,ZhanKnox2004,Jing2006,Rudd2008,Guillet2010,Casarini2011,vanDaalen2011} and, subsequently, weak lensing measurements \citep[e.g.][]{Semboloni2011,Semboloni2013,Yang2013,Zentner2013}, the strong lensing properties of clusters \citep[e.g.][]{Mead2010,Killedar2012}, and the halo mass function \citep[e.g.][]{Stanek2009,Cui2012,Sawala2013,Balaguera-Antolinez2013,Martizzi2013,Velliscig2014}. To complicate matters further, different authors studying the same aspects of galaxy formation often find different and sometimes even contradictory results, depending not only on which physical processes are modelled but also on the choice of numerical code, and particularly on the implementation of subgrid recipes for feedback from star formation and Active Galactic Nuclei (hereafter AGN) \citep[e.g.][]{Scannapieco2012}. Until a consensus can be reached, it is therefore important to determine the range of values that observables can take depending on whether certain baryonic processes are included in a model, and the way in which they are implemented.

In this paper, we aim to quantify the effects of baryons and galaxy formation on the two-point real-space correlation function. Specifically, we will investigate how the redshift zero galaxy and subhalo correlation functions and the galaxy-matter cross correlation, which is observable through galaxy-galaxy lensing, are changed if baryonic processes are allowed to influence the distribution of matter to varying degrees, i.e.\ using different feedback models. To this end, we will use the reference and AGN models from the OverWhelmingly Large Simulations project \citep[OWLS,][]{Schaye2010}. These were also employed by \citet{vanDaalen2011} and we have since repeated them using larger volumes, more particles and a more up-to-date cosmology. The AGN model is particularly relevant, as it has been shown to reproduce many relevant X-ray and optical observations of groups and clusters \citep{McCarthy2010, McCarthy2011,Stott2012}.

Any changes in the clustering of objects brought about by galaxy formation can enter into the correlation function in two ways. The first and most well-established effect is due to a change in the mass of the objects. For example, assuming that higher-mass haloes are more strongly clustered, if supernova feedback systematically lowers the stellar content of haloes, then a model which includes this process is expected to show increased clustering at fixed stellar mass relative to one that does not.\footnote{Situations in which feedback would have the reverse effect are possible in principle. For example, if the stellar mass - halo mass relation were flat where AGN feedback is important and had a large scatter, then the stellar mass of some galaxies inhabiting such haloes could be lower than that of galaxies in lower-mass haloes. As a result, the most massive galaxies would reside in intermediate mass haloes. However, such a scenario is not supported by our simulations.} Likewise, the clustering of haloes at fixed halo mass is also expected to show increased clustering when efficient feedback is included, due to the total mass of the halo being lowered. Secondly, the positions of galaxies and haloes may shift due to changes in the physics: if the mass within a certain radius around an object changes, then the gravitational force acting on those scales will change as well, affecting the dynamics of nearby galaxies and haloes. Moreover, tidal stripping, and hence also dynamical friction, will affect satellites differently if baryonic processes change the density profiles of either the satellites or the host haloes. We will consider both types of effects here; most importantly, we will disentangle the two and show what effects remain after we account for the change in halo mass, as could be done approximately by selecting samples with constant number density. As we will see, not all shifts in position average out, nor can the modification of the halo profiles be ignored.

Quantifying the significance of the various ways in which clustering measurements may deviate from those in a dark matter only universe is vital for the improvement of current models employed in clustering studies. Typically these are based on the distribution of dark matter alone, be they semi-analytical models \citep[see][ for a review]{Baugh2006}, a combination of halo occupation distribution (HOD) and halo models \citep[e.g.][]{JingMoBoerner1998,BerlindWeinberg2002,CooraySheth2002,YangMovandenBosch2003,Kravtsov2004,Tinker2005,Wechsler2006,vandenBosch2013} or subhalo abundance matching (SHAM) models \citep[e.g.][]{ValeOstriker2004,Shankar2006,Conroy2006,Moster2010,Guo2010,Behroozi2010,Simha2013}. It is therefore important to investigate which ingredients may currently be missing from such efforts.

The effects of galaxy formation on subhalo-subhalo clustering were previously considered by \citet{Weinberg2008} and \citet{Simha2012}. \citet{Weinberg2008} compared the clustering of objects at fixed number density in a dark matter only simulation with a baryonic simulation including weak supernova feedback but no feedback from AGN, and with identical initial conditions. They found that subhaloes cluster more strongly on small scales in the baryonic simulation due to the increased survival rate of baryonic satellites during infall. While we find a similar increase in the autocorrelation of subhaloes on small scales ($r \la r_\mathrm{vir}$) -- with a corresponding decrease in clustering on slightly larger scales -- we point out that such results may be biased, due to the difficulties of detecting infalling dark matter satellites \citep[e.g.][, see our Appendix~\ref{sec:linkedfractions}]{Muldrew2011}.

\citet{Simha2012} extended the work of \citet{Weinberg2008} in several ways, among which are the addition of more effective stellar feedback and the use of the mass of the subhalo at infall, rather than the current mass, when assigning galaxy properties to the subhaloes. They find that the addition of effective feedback causes the discrepancies between clustering in hydrodynamical simulations and results from subhalo abundance matching to increase. They demonstrate that the two-point correlation function of baryonic subhaloes can be recovered to better than $15\%$ on scales $r>2\runit$ when winds are included, but that the discrepancy at smaller scales in these simulations can be up to a factor of a few. The galaxy correlation function is reproduced much better if the stellar mass threshold is raised; however, as these simulations do not contain any form of feedback that is effective at high stellar masses, we would expect the further addition of a process like AGN feedback to exacerbate the discrepancy between subhalo abundance matching results and hydrodynamical simulations for massive galaxies.

This paper is organized as follows. We will briefly introduce our simulations and explain how we calculate the relevant quantities in \S\ref{sec:methods}. Here we will also discuss how we identify the same halo in different simulations, an essential step in order to separate the change in halo mass from other effects. We present our results in \S\ref{sec:results} and summarise our findings in \S\ref{sec:summary}. Finally, we show the convergence with resolution and box size in Appendix~\ref{sec:convergence} and consider the fraction of subhaloes successfully linked between simulation in Appendix~\ref{sec:linkedfractions}.

\section{Method}
\label{sec:methods}
\subsection{Simulations}
\label{subsec:simulations}
We consider three models from the OWLS project \citep{Schaye2010}: \textit{DMONLY}, \textit{REF} and \textit{AGN}. All of these simulations were run with a modified version of \textsc{gadget iii}, the smoothed-particle hydrodynamics (SPH) code last described in \citet{Springel2005c}. We will discuss the models employed briefly below.

In order to study relatively low-mass objects while also simulating a volume that is sufficiently large to obtain a statistical sample of high-mass objects, we combine the results of simulations with different box sizes. For each model, we ran simulations in periodic boxes of comoving side lengths $L=200$ and $400\runit$, both with $N^3=1024^3$ CDM particles and -- with the exception of \textit{DMONLY} -- an equal number of baryonic particles. The gravitational forces are softened on a comoving scale of $1/25$ of the initial mean inter-particle spacing, $L/N$, but the softening length is limited to a maximum physical scale of $1\runitk [L/(100\runit)]$. The particle masses in the baryonic \textit{L200} (\textit{L400}) simulations are $4.68\times 10^8\munit$ ($3.75\times 10^9\munit$) for dark matter and $9.41\times 10^7\munit$ ($7.53\times 10^8\munit$) for the baryons. We will use the higher-resolution \textit{L200} simulations to study the clustering of galaxies with stellar mass $M_*<10^{11}\munit$ and subhaloes with total mass $M_\mathrm{sh}<10^{13}\munit$, while taking advantage of the larger volume of the \textit{L400} simulations to study higher masses. When considering cross-correlations with the matter distribution, resolution is more important than volume, and we use the \textit{L200} simulations at all masses. We discuss our choice of mass limits in Appendix~\ref{sec:convergence}, where we also show resolution tests. All the simulations we employ in this paper were run with a set of cosmological parameters derived from the Wilkinson Microwave Anisotropy Probe (WMAP) 7-year results \citep{Komatsu2011}, given by \{$\Omega_\mathrm{m}$, $\Omega_\mathrm{b}$, $\Omega_\mathrm{\Lambda}$, $\sigma_8$, $n_\mathrm{s}$, $h$\} = \{$0.272$, $0.0455$, $0.728$, $0.81$, $0.967$, $0.704$\}. It is important to note that all simulations with identical box sizes were run with identical initial conditions, which allows us to compare the effects of baryons and galaxy formation for the exact same objects.

The \textit{DMONLY} simulation, as its name suggests, contains only dark matter. This provides us with a useful baseline model for testing the impact of baryon physics.

The \textit{REF} simulation is the reference OWLS model. It includes sub-grid recipes for star formation \citep{SchayeDallaVecchia2008}, radiative (metal-line) cooling and heating \citep{Wiersma2009a}, stellar evolution, mass loss from massive stars and chemical enrichment \citep{Wiersma2009b} and a kinetic prescription for supernova feedback \citep{DallaVecchiaSchaye2008}. The reference simulation is not intended to be the most realistic, but instead includes only those physical processes most typically found in simulations of galaxy formation.

The third and final simulation we consider here, \textit{AGN}, adds feedback from accreting supermassive black holes to the reference simulation. AGN feedback was modelled following the prescription of \citet{BoothSchaye2009}, which built on the model of \citet{Springel2005b}. We believe \textit{AGN} to be our most realistic model, as it is the only model that solves the well-known overcooling problem \citep[e.g.][]{Balogh2001} and that reproduces the observed properties of groups \citep{McCarthy2010, McCarthy2011,Stott2012}. Specifically, this model has been shown to reproduce the gas density, temperature, entropy, and metallicity profiles inferred from X-ray observations, as well as the stellar masses, star formation rates, and stellar age distributions inferred from optical observations of low-redshift groups of galaxies. \citet{vanDaalen2011} used this model to show that AGN feedback has a dramatic effect on the clustering of matter; here we wish to investigate whether the effect on the clustering of galaxies and subhaloes is equally important.

\subsection{Calculating correlation functions}
\label{subsec:calculating}
The correlation function, $\xi(r)$, returns the excess probability, relative to a random distribution, of finding two objects at a given separation $r$. It is therefore a measure of the clustering of these objects as a function of scale. As our simulations contain only a moderate number of resolved objects (i.e.\ galaxies and (sub)haloes), we do not need to resort to approximations that are common in the calculation of two-point clustering statistics. Instead, we can use a parallelised brute force approach in which we obtain the (cross-)correlation function through simple pair counts, using the relation:
\begin{equation}
\label{correl}
\xi_{XY}(r)=\frac{DD_{XY}(r)}{RR_{XY}(r)}-1.
\end{equation}
Here $X$ and $Y$ denote two (not necessarily distinct) sets of objects (e.g.\ galaxies and particles or galaxies and galaxies), $DD_{XY}(r)$ is the number of unique pairs consisting of an object from set $X$ and an object from set $Y$ separated by a distance $r$, and $RR_{XY}(r)$ is the expected number of pairs at this separation if the positions of the objects in these sets were random. As our simulations are carried out with periodic boundary conditions, more complicated expressions involving cross terms of the form $DR_{XY}(r)$ \citep[e.g.][]{LandySzalay1993} are not necessary, nor do we need to actually create random fields; instead, we can simply compute the term in the denominator analytically.

The basic functions that we will consider in this paper are the galaxy autocorrelation function, $\xi_\mathrm{gg}$, the galaxy-mass cross correlation function, $\xi_\mathrm{gm}$, the subhalo autocorrelation function, $\xi_\mathrm{ss}$, and the subhalo-mass cross correlation function, $\xi_\mathrm{sm}$. We divide galaxies and subhaloes into different bins according to their stellar and subhalo dark matter mass, respectively. When cross-correlating with matter, we weight particles by their mass. To keep the computation time manageable, we use only $25\%$ of all particles for the lowest mass bin of the simulations with $(2\times)1024^3$ particles, randomly selected. In all other mass bins, we cross-correlate with the full particle distribution. We have verified that this does not influence our results in any way. Throughout this paper we will focus on the three-dimensional correlation function. We will only show the correlation functions in radial bins where the number of pairs exceeds $10$, to prevent our results from being dominated by spurious clumping. We take the position of our objects to be the position of their most-bound particle, and assign each galaxy a mass equal to the total mass in stars in its subhalo. Finally, we confine our analysis to scales $r \la 20\runit$, corresponding to at most $1/10$th of box size, in order to avoid the effects of missing large-scale modes.

\subsection{Linking haloes between different simulations}
\label{subsec:linking}
As discussed previously, there are two main ways in which the two-point correlation function may be affected by baryonic processes: through changes in the masses of objects, and through shifts in their positions. To disentangle the two effects, we make use of the fact that all OWLS models were run from identical initial conditions, allowing us to identify the same objects in different simulations. In this way we can assign each object in simulation B the mass that the same object possesses in simulation A, thereby isolating the effect of changes in the positions of objects on the clustering signal.

Haloes are identified in our simulations using the Friends-of-Friends algorithm (run on the dark matter particles, with linking length $0.2$) combined with a spherical overdensity finder, as implemented in the \textsc{subfind} algorithm \citep{Springel2001,Dolag2009}. For every (sub)halo in simulation A we flag the $N_\mathrm{mb}$ most-bound dark matter particles, meaning the particles with the highest absolute binding energy. Next, we locate these particles in the other simulations, using the unique number associated with every particle. If we find a (sub)halo in simulation B that contains at least $50\%$ of these flagged particles, a first link is made. The link is confirmed if, by repeating the process starting from simulation B, the previous (sub)halo in simulation A is found.

Here we use $N_\mathrm{mb}=50$, but we have verified that our results are insensitive to this choice \citep[see][]{Velliscig2014}. For haloes with less than $N_\mathrm{mb}$ dark matter particles, all dark matter particles are used. The fraction of haloes linked quickly increases as a function of mass, reaching essentially unity for sufficiently well-resolved haloes. For all subhaloes employed in this work, the linked fraction of \textit{DMONLY} subhaloes typically exceeds $99\%$, the exception being the lowest mass bin where the linked fraction is around $98\%$. However, at small separations the linked fraction can be much smaller. This is explored in more detail in Appendix~\ref{sec:linkedfractions}.

\section{Results}
\label{sec:results}
In this section we will explore the effects of baryon physics on the two-point correlation function at redshift zero. We will first consider the galaxy-galaxy and galaxy-matter correlation functions as these are the most directly observable. Since stellar masses are strongly model-dependent, we will switch from galaxies to subhaloes in \S\ref{subsec:subhaloes}, which allows us to examine how clustering statistics derived from dark matter only simulations will differ from those including baryons. Finally, in \S\ref{subsec:withlinking}, we will take the change in the mass of subhaloes out of the equation, and consider the change in the correlation function for the exact same objects as a function of the model used.

\begin{figure}
\begin{center}
\includegraphics[width=1.0\columnwidth, trim=0mm -14mm 5mm -3mm]{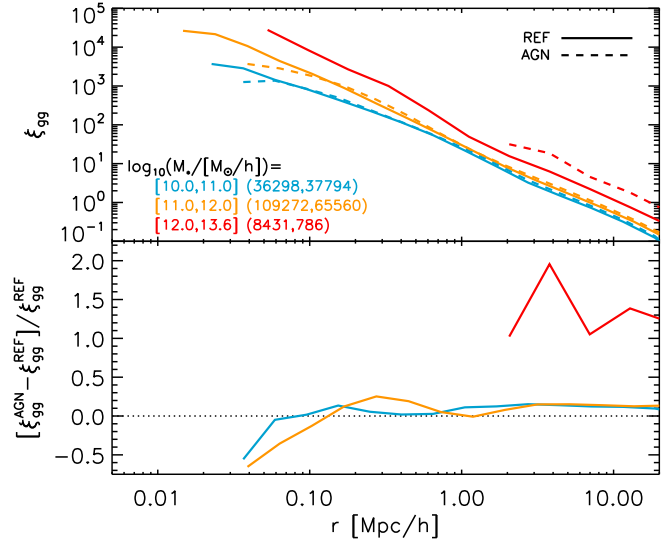}
\caption{The galaxy autocorrelation function for the \textit{REF} and \textit{AGN} simulations (top), as well as the fractional difference between the two (bottom). Different colours correspond to different stellar masses, as indicated in the legend. The legend also shows the number of galaxies in each bin for each simulation (\textit{REF},\textit{AGN}). At any mass, galaxies in \textit{AGN} are more highly clustered than those in \textit{REF} on large scales, an effect that increases sharply above $10^{12}\munit$, where AGN feedback is most important. Note that these effects may be underestimated for the two highest mass bins for reasons discussed in \S\ref{subsubsec:caveats}. The relative decrease in clustering for the \textit{AGN} simulation on small scales is mostly a numerical effect (see text).}
\label{gal-gal}
\end{center}
\end{figure}

\subsection{Clustering of galaxies}
\label{subsec:galaxies}
\subsubsection{Autocorrelation}
\label{subsubsec:galaxiesauto}
In Figure~\ref{gal-gal} we plot the galaxy autocorrelation functions, $\xi_\mathrm{gg}(r)$, for models \textit{REF} and \textit{AGN} in three different bins of stellar mass, as indicated in the legend. The bottom panel shows the relative difference in the clustering strength of galaxies in these models. Since the clustering of haloes increases with mass, and since AGN feedback reduces the stellar content of massive haloes, one would expect galaxies in the \textit{AGN} simulation to be more strongly clustered at fixed (high) stellar mass. As higher-mass galaxies are expected to host more powerful AGN, this effect is expected to increase with mass. This is indeed what we observe in Figure~\ref{gal-gal}: as long as we consider sufficiently large scales, galaxies in the \textit{AGN} simulation show increased clustering relative to those in \textit{REF}, and the relative difference between clustering strengths in the two simulations tends to increase with mass. For galaxies with stellar masses $M_*<10^{10}\munit$ we expect the effect to be minor, since in such low-mass objects feedback is controlled by stellar rather than AGN feedback in these models \citep[e.g.][]{Haas2013}.

Also indicated in the legend are the number of galaxies in each mass bin for each simulation, the first number corresponding to \textit{REF} and the second to \textit{AGN}. Because AGN feedback systematically lowers the stellar content of massive haloes, and since the number density of haloes decreases with mass, the \textit{AGN} simulation suffers from somewhat worse statistics at high stellar masses than the \textit{REF} simulation. However, this effect is only seen in the highest mass bin, $M_*>10^{12}\munit$, and even in this mass range we can still draw robust conclusions for scales $r>2\runit$.

Note that any two subhaloes must have a finite minimum distance between them in order to, on the one hand, be recognised as separate objects, and on the other, not be tidally destroyed. As we identify galaxies by the subhaloes they occupy, this causes a slight turnover in the galaxy correlation functions on small scales. Since this minimum distance increases with the size and therefore mass of the subhaloes hosting the galaxies, at fixed stellar mass this turnover is seen at larger scales in the model \textit{AGN} than in \textit{REF}. This in turn causes the galaxies in \textit{AGN} to appear less clustered on small scales.

\subsubsection{Cross-correlation with matter}
\label{subsubsec:galaxiescross}
\begin{figure}
\begin{center}
\includegraphics[width=1.0\columnwidth, trim=0mm -14mm 5mm -3mm]{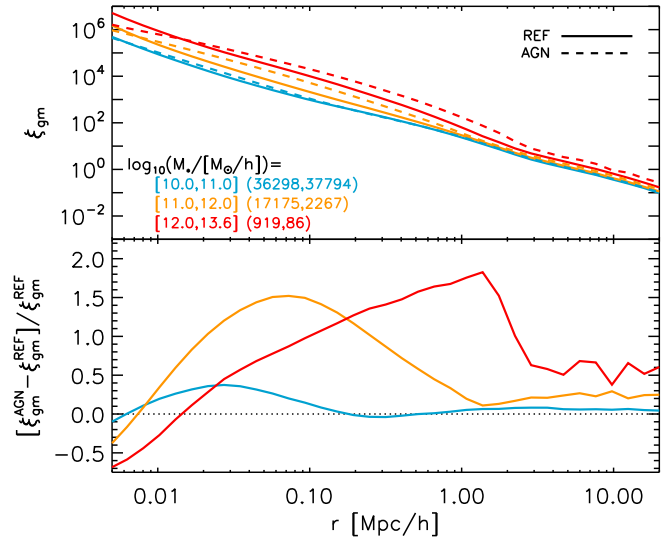}
\caption{As in Figure~\ref{gal-gal}, but now showing the galaxy-matter cross-correlation function for the \textit{REF} and \textit{AGN} simulations. Except for sub-galactic scales, AGN feedback tends to increase the clustering of galaxies with matter at fixed stellar mass. Both the overall magnitude of the effect and the length scales over which it occurs increase with stellar mass, and for $M_*>10^{12}\munit$ the increase in clustering with the matter distribution reaches values as high as $180\%$.}
\label{gal-mass}
\end{center}
\end{figure}

Figure~\ref{gal-mass} shows the galaxy-matter cross-correlation functions for these simulations, which are relevant for galaxy-galaxy lensing. Due to the high number of particles relative to the number of galaxies, the statistics are significantly improved relative to Figure~\ref{gal-gal}, and we can see clearly that including AGN feedback greatly increases the clustering of matter and galaxies at fixed stellar mass.\footnote{Note that the number of objects in the two most massive bins, shown in the legend, is lower than for the autocorrelation function. This is because we now use the higher-resolution \textit{L200} for all mass bins, whereas we previously used \textit{L400} for the two highest mass bins to obtain better statistics (see \S\ref{subsec:simulations}).}. The relative increase of clustering with mass is more strongly scale-dependent than for the galaxy-galaxy case. The relative difference in clustering strength between \textit{AGN} and \textit{REF} is largest around $1\runit$ for the most massive galaxies, where galaxies at fixed stellar mass are nearly twice as strongly clustered with matter when AGN are included. At larger scales, \textit{AGN} always shows $\sim 50\%$ stronger clustering than \textit{REF} for $M_*>10^{12}\munit$. Even for galaxies in the stellar mass range $10^{11}<M_*/[\mathrm{M}_{\sun}/h]<10^{12}$ we see an increase in clustering of up to $150\%$ around $70\runitk$, and an offset of $\sim 20\%$ at all larger scales.

Interestingly, the relative difference in the galaxy-matter cross-correlation functions between \textit{AGN} and \textit{REF} increases towards smaller scales before suddenly dropping, causing galaxies to become \emph{less} strongly clustered with the matter distribution in the \textit{AGN} simulation on the very smallest scales probed here. This behaviour is caused by two competing effects, a point we will return to when discussing the subhalo-matter cross-correlation function in the next sections. On the one hand, the lowering of the stellar mass by AGN feedback tends to increase clustering at fixed stellar mass, and more so towards smaller scales, as galaxies of the same stellar mass now inhabit denser environments. On the other hand, as shown in e.g.\ \citet{Velliscig2014}, a large amount of gas -- and even dark matter -- is removed from the galaxy, and sometimes from the halo entirely, decreasing the density peaks in the matter distribution \citep[see e.g.][]{vanDaalen2011}. As we can see from Figure~\ref{gal-mass}, the latter effect dominates on sub-galaxy scales ($r \la 10\runitk$).

\subsubsection{Caveats}
\label{subsubsec:caveats}
We note that the effect of AGN feedback may be underestimated for massive galaxies due to two effects. The first only applies to the two highest mass bins and only to results based on the \textit{L400} runs (i.e.\ the autocorrelation functions): the implementation of AGN feedback in these simulations is somewhat resolution dependent, and as a consequence its effect is weaker in the $400\runit$ box than in the $200\runit$ simulation. This is because the seed black holes can only be injected into resolved haloes, which corresponds to a minimum mass, that is $8$ times higher in the \textit{L400} simulation than in the \textit{L200} simulation (i.e.\ the difference in mass resolution). The result is that AGN feedback in the $400\runit$ box, used in the two highest mass bins in Figures \ref{gal-gal} and \ref{gal-mass}, may be too weak for galaxies occupying haloes with masses $M \la 10^{13}\munit$. In fact, while the effect of resolution is small for galaxies with masses $M_*>10^{12}\munit$, for $10^{11}<M_*/[\mathrm{M}_{\sun}/h]<10^{12}$ the effect is significant: when using the higher-resolution \textit{L200} simulation in this mass bin, we find an increase in galaxy-galaxy clustering relative to \textit{REF} of $\sim 50\%$ for $r \ga 2\runit$.

The second effect is due to the way stellar mass is estimated in observations, where the use of an aperture excludes intra\-cluster light. For the more massive galaxies in our sample, which host the most powerful AGN, this aperture size is typically significantly smaller than the size of the region containing the stars. However, simulated galaxies are assigned a stellar mass equal to the total mass in stars in its subhalo. The stellar mass of our most massive galaxies is therefore significantly higher than would be estimated observationally. Hence, the strong effects of AGN feedback that we find will be relevant for lower \emph{observed} stellar masses than suggested by our plots.

Regardless, even without taking these effects into account, it is clear that AGN feedback plays an important role in the clustering of galaxies and matter, and should not be ignored in theoretical models that aim to predict $\xi_\mathrm{gm}(r)$ to $\sim 10\%$ accuracy or better, even when only considering relatively low stellar masses ($M_*=10^{10}-10^{11}\munit$).

At this point it is important to note that although our model \textit{AGN} reproduces the stellar masses of group-sized haloes relatively well \citep{McCarthy2010,McCarthy2011}, predicted stellar masses are generally strongly model-dependent, as well as cosmology-dependent. Abundance matching studies, on the other hand, reproduce the stellar mass-halo mass relation by construction \citep[e.g.][]{Moster2010}. Since clustering models typically employ the results from such studies, which in turn rely on dark matter only simulations, it is useful to consider the clustering of the subhaloes that host the galaxies and to select objects by their total subhalo mass, instead of by their stellar mass. This also allows us to consider the effect of galaxy formation relative to a dark matter only scenario. For the remainder of this paper, we will therefore focus on the clustering of subhaloes.

\subsection{Clustering of subhaloes}
\label{subsec:subhaloes}
\subsubsection{Autocorrelation}
\label{subsubsec:subhaloesauto}
\begin{figure}
\begin{center}
\includegraphics[width=1.0\columnwidth, trim=46mm -9mm 17mm -2mm]{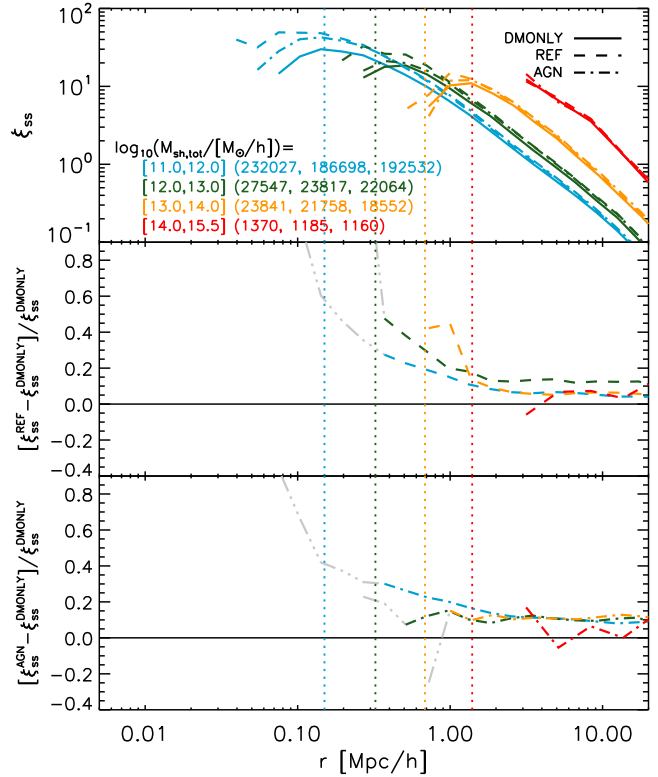}
\caption{The subhalo autocorrelation function, $\xi_\mathrm{ss}(r)$, for \textit{DMONLY} (solid), \textit{REF} (dashed) and \textit{AGN} (dot-dashed lines), and the fractional differences between them. Different colours are used for different total subhalo masses, and the number of objects in each bin is indicated in the legend (\textit{DMONLY}, \textit{REF}, \textit{AGN}). \textit{Top:} The correlation functions for the three simulations. Vertical dotted lines indicate the median $r_\mathrm{vir}$ of the subhaloes. \textit{Middle:} The fractional difference of subhalo clustering in \textit{REF} relative to \textit{DMONLY}. The curves are greyed out for radii where they may be biased due to subhalo non-detections (see Appendix~\ref{sec:linkedfractions}). \textit{Bottom:} The fractional difference of subhalo clustering in \textit{AGN} relative to \textit{DMONLY}. Both baryonic simulations show increased clustering, and this effect is stronger on smaller scales. Note that the range on the y-axis is much smaller here than in Figure~\ref{gal-gal}.}
\label{sub-sub}
\end{center}
\end{figure}

The top panel of Figure~\ref{sub-sub} shows the subhalo autocorrelation function, $\xi_\mathrm{ss}(r)$, for three different simulations: \textit{DMONLY}, \textit{REF} and \textit{AGN}. Different colours indicate different subsamples, selected by the \emph{total} mass of the subhaloes, $M_\mathrm{sh,tot}$, though we note that the results would have been very similar had we selected by dark matter mass. The correlation functions are displayed in the top panel, while in the middle panel and bottom panels the baryonic simulations are compared to \textit{DMONLY}. From the top panel we can already see that subhalo clustering in the dark matter only simulation behaves quite differently from that in the baryonic models, especially on small scales ($r \la 1\runit$). Vertical dotted lines indicate the median virial radii\footnote{We computed a characteristic size, $r_\mathrm{vir}$, for each subhalo by taking its total mass, $M_\mathrm{sh,tot}$, and treating it as the mass within a region with a mean overdensity of $\Delta=200$ relative to $\rho_\mathrm{crit}$ (i.e.\ $r_\mathrm{vir} \approx r_\mathrm{200c}$).  For reference, for a typical dark matter halo $r_\mathrm{500c} \sim 0.65-0.75\,r_\mathrm{200c}$, where $r_\mathrm{500c}$ corresponds to the radius out to which the dominant baryonic component (hot gas) of groups and clusters is typically measured \citep[e.g.][]{Vikhlinin2006}.} of subhaloes in each mass bin, which are similar to the scale at which the subhalo correlation functions for \textit{DMONLY} turn over.

At the high-mass end, all three simulations show very similar behaviour. Looking at the middle and bottom panels, where we compare the autocorrelation of subhaloes in \textit{REF} and \textit{AGN} respectively to that in \textit{DMONLY}, we see that all subhaloes in the baryonic simulations are typically $\sim 10\%$ more strongly clustered on large scales than their dark matter only counterparts. As we will demonstrate in \S\ref{subsec:withlinking}, this difference is due to the reduction of subhalo mass caused by baryonic processes. For the larger subhaloes, $10^{13}<M_\mathrm{sh,tot}/[\mathrm{M}_{\sun}/h]<10^{14}$, this offset is somewhat larger when AGN feedback is included, because supernova feedback alone cannot change the subhalo mass by as much as it can for lower halo masses \citep[e.g.][]{Sawala2013,Velliscig2014}. The offset in clustering strength relative to \textit{DMONLY} of the lowest-mass subhaloes is also slightly increased by the addition of AGN: while the masses of these subhaloes may seem to be somewhat low to be significantly affected by AGN feedback, we should keep in mind that satellite subhaloes may have lost part of their mass through tidal stripping. Moreover, these would correspond to subhaloes of a higher mass in a \textit{DMONLY} simulation, as a significant fraction of the mass has been expelled. Additionally, low-mass subhaloes do not need to host AGN themselves to be affected by them: satellites in groups and clusters are sensitive to changes in the host halo profile and possibly increased stripping caused by the powerful AGN in the more massive galaxies in their environment.

The differences between the baryonic and dark matter only simulations increase rapidly for $r<2r_\mathrm{vir}$, at least for $M_\mathrm{sh,tot}<10^{14}\munit$. As we can see most easily in the top panel, subhaloes in the \textit{REF} simulation are significantly more clustered on small scales than those in the \textit{AGN} simulation, which seems to contradict the results of the previous section. This is because subhaloes in the \textit{REF} simulation are more compact at fixed mass than those in the \textit{AGN} simulation, due to the additional form of feedback in the latter which removes more material from the centre and lowers the concentration in the inner parts of the subhaloes. However, the haloes in the \textit{AGN} simulation are still more compact than those in \textit{DMONLY} \citep[see e.g.][]{Velliscig2014}. The increased concentration of subhaloes in baryonic simulations allows them to be identified as separate objects down to smaller scales, and also to withstand the effects of tidal stripping longer than their dark matter only counterparts. Both these effects tend to increase the clustering on small scales. This relative increase in the number density of subhaloes close to the centres of haloes in baryonic simulations was seen before by e.g.\ \citet{Maccio2006}, \citet{Libeskind2010}, \citet{Romano-Diaz2010} and \citet{SchewtschenkoMaccio2011} (although \citealt{Romano-Diaz2010} note that without strong feedback, the effect may be reversed). On the other hand, baryonic subhaloes are generally less massive when they are centrals, and those that become satellites typically fall in later due to the smaller virial radius of the main halo compared to a pure dark matter run, which means that they should experience less dynamical friction on scales where tidal stripping is not yet important. This is indeed what \citet{SchewtschenkoMaccio2011} find, although this effect cannot be seen for the mass-selected sample shown in Figure~\ref{sub-sub} due to the much larger effect of the change in mass.

We explore the clustering behaviour of baryonic satellites in more detail in \S\ref{subsubsec:linkedauto}. For now, we note that if our ability to detect baryonic subhaloes down to smaller radii than pure dark matter ones were the dominant cause of an increased number density of subhaloes at small separations in \textit{REF} and \textit{AGN}, this would introduce a bias towards observing a stronger clustering signal in baryonic models on scales $r \la 2r_\mathrm{vir}$.\footnote{We thank Raul Angulo for pointing out this potential problem.} We discuss this possible source of error in Appendix~\ref{sec:linkedfractions}, and based on the results reported there we have chosen to show the relative differences in clustering as grey dot-dot-dot-dashed curves in Figure~\ref{sub-sub} for subhalo masses and scales that may be significantly affected by this bias.

Comparing Figures~\ref{gal-gal} and \ref{sub-sub}, we see that the single act of adding AGN feedback affects the clustering of galaxies and subhaloes very differently. For galaxies, a strong increase in clustering is found for the highest-mass galaxies, and on large scales, since the same subhaloes host galaxies with a much lower stellar mass when AGN feedback is added. Low-mass galaxies are, however, not strongly affected by AGN feedback. For subhaloes, on the other hand, we find that the largest effects are found on small scales, and especially at the lowest masses: we find a strong \emph{decrease} in clustering for $r \la r_\mathrm{vir}$ when adding AGN feedback to the reference model, regardless of halo mass, and far less change on large scales. These two main differences have two different causes. The large-scale differences between the effect of AGN feedback on galaxies and on subhaloes is that while AGN are powerful enough to quench star formation and to remove a lot of gas from galaxies, thus lowering the stellar mass, they are not powerful enough to significantly change the halo mass. However, as is shown in detail by \citet{Velliscig2014}, and as we will also see in the next section, they do have a significant effect on the density profiles of subhaloes, and through this on their distribution. At fixed mass, the subhaloes in \textit{REF} are more compact and more massive than those in \textit{AGN}, causing both the satellite survival rate and the dynamical friction experienced by satellites to increase, which in turn causes the small-scale differences in clustering we just discussed.

\begin{figure}
\begin{center}
\includegraphics[width=1.0\columnwidth, trim=46mm -9mm 17mm -2mm]{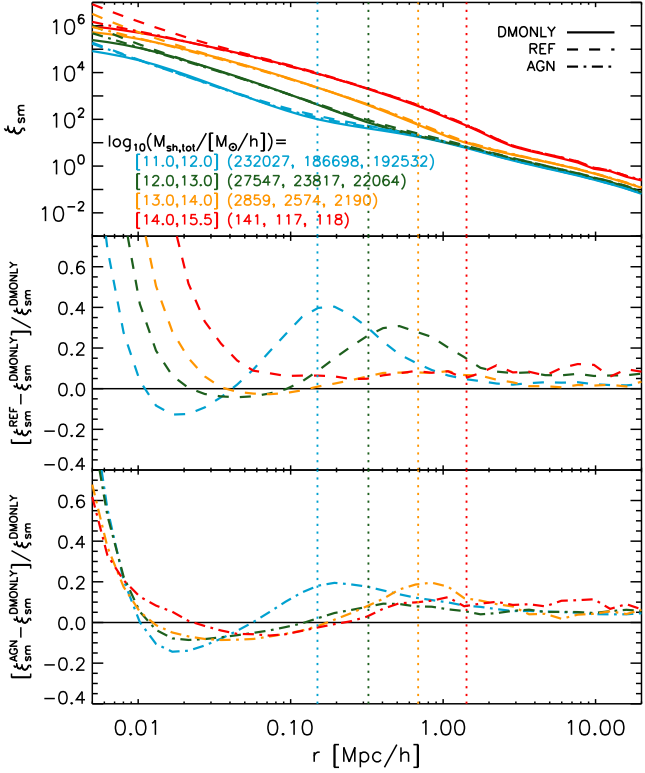}
\caption{As Figure~\ref{sub-sub}, but now for the subhalo-mass cross-correlation function, $\xi_\mathrm{sm}(r)$. Subhaloes are generally more strongly clustered with matter in the baryonic simulations than in \textit{DMONLY}. The largest differences are found for \textit{REF}, for which $\xi_\mathrm{sm}(r)$ can be up to $40\%$ higher on intermediate scales for the lowest-mass subhaloes, and much higher still for any subhalo mass if sufficiently small scales are considered. There is also a constant $5\%$ difference in favour of the baryonic simulations on large scales, regardless of subhalo mass. While the \textit{AGN} model seems to increase clustering at fixed subhalo mass less than \textit{REF}, it does show a stronger \emph{decrease} in clustering up to scales $r \sim 10^2\runitk$. Note that in both cases the clustering differences between the models are strongly non-monotonic, which is caused by the interplay between the change in the total subhalo mass and the change in the subhalo mass profiles.}
\label{sub-mass}
\end{center}
\end{figure}

\subsubsection{Cross-correlation with matter}
\label{subsubsec:subhaloescross}
We consider the subhalo-mass cross correlation function in Figure~\ref{sub-mass}. From the top and middle panels we observe, as was the case for galaxy-galaxy clustering, that on the smallest scales and at fixed total mass, subhaloes cluster far more strongly with matter in the baryonic simulations than in the dark matter only simulations. Additionally, there is a constant $5\%$ offset in favour of baryonic simulations on the largest scales, for all halo masses. The baryonic bias increases as we move from large scales towards the virial radius, but, interestingly, the strength of the effect decreases below scales approximately corresponding to $r_\mathrm{vir}$ before picking up again at the smallest scales shown. This decrease below $r_\mathrm{vir}$ even causes the lowest-mass \textit{DMONLY} subhaloes to be more strongly clustered than their \textit{REF} counterparts around $r=20\runitk$. For \textit{AGN}, this happens even for the highest-mass subhaloes, and over a larger range of scales.

As we will show in the next section, the strongly non-monotonic behaviour of the relative difference in $\xi_\mathrm{sm}$ between the baryonic simulations and \textit{DMONLY} is caused by two counteracting effects. On the one hand, the lowered halo masses in the baryonic simulations tend to increase clustering at fixed mass on all scales. On the other hand, while the dissipation associated with galaxy formation causes the inner halo profile to steepen, increasing clustering on small scales, the associated feedback causes the outer layers of the halo to expand, decreasing clustering on intermediate scales. This effect is stronger when AGN feedback is included. Note that we observe similar behaviour for the relative differences between the galaxy-matter cross-correlation functions for \textit{REF} and \textit{AGN}.

Furthermore, by comparing the bottom two panels, we can see that for low halo masses ($M_\mathrm{sh,tot}<10^{12}\munit$), for which AGN feedback is not very important, the small-scale clustering of haloes in \textit{REF} and \textit{AGN} is nearly identical, while subhaloes and matter cluster much more weakly on a range of scales around $r_\mathrm{vir}$ in \textit{AGN}. On the other hand, for higher-mass haloes ($M_\mathrm{sh,tot}>10^{12}\munit$), significant differences can be seen down from the smallest scales out to $r \sim 1\runit$. This again confirms the strong effect that AGN feedback has on the mass distribution: the higher the mass of the halo, the more important feedback from supermassive black holes is in removing material from the centre. This in turn flattens the mass profiles of the haloes and smooths out the density peaks, decreasing the small-scale lensing signal relative to \textit{REF}.

As we have already pointed out several times, the most important cause of the increase in clustering due to galaxy formation with strong feedback is the lowering of the mass of objects. However, secondary effects, such as the resulting changes in the dynamics and density profiles of haloes, are also expected to be significant. To disentangle these types of effects, we will use our linking scheme to match subhaloes between different simulations, allowing us to see if any significant difference in the clustering remains once the change in mass has been accounted for.

\subsection{Accounting for the change in mass}
\label{subsec:withlinking}
\begin{figure}
\begin{center}
\includegraphics[width=1.0\columnwidth, trim=46mm -9mm 17mm -2mm]{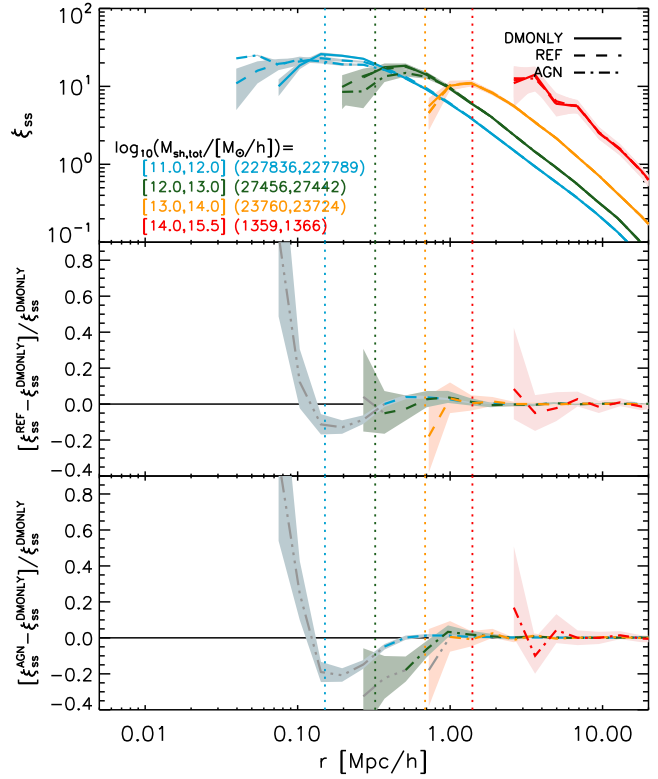}
\caption{As Figure~\ref{sub-sub}, but now only showing the autocorrelation functions for subhaloes linked between a baryonic simulation and \textit{DMONLY}, and selected based on their mass in the latter. Relative to Figure~\ref{sub-sub}, this procedure removes the effects of changes in the subhalo masses. As the numbers in the legend imply, almost the exact same haloes are linked with dark matter only haloes in both cases. The bottom two panels immediately show that in all cases no differences $\ga 5\%$ in $\xi_\mathrm{ss}$ remain on scales $r \gg r_\mathrm{vir}$, indicating that the differences we saw in Figure~\ref{sub-sub} on these scales were due to the masses of the objects changing. For smaller scales, and especially for low-mass subhaloes, the change in dynamics of the objects in the baryonic simulations can have significant effects, which can primarily be seen as a decrease in clustering on scales $r \la 2r_\mathrm{vir}$. Shaded areas indicate the regions allowed by 1$\sigma$ bootstrap errors, which show that the relative small-scale decrease of clustering of low-mass baryonic subhaloes is significant.}
\label{sub-sub_linked}
\end{center}
\end{figure}

\begin{figure}
\begin{center}
\includegraphics[width=1.0\columnwidth, trim=46mm -9mm 17mm -2mm]{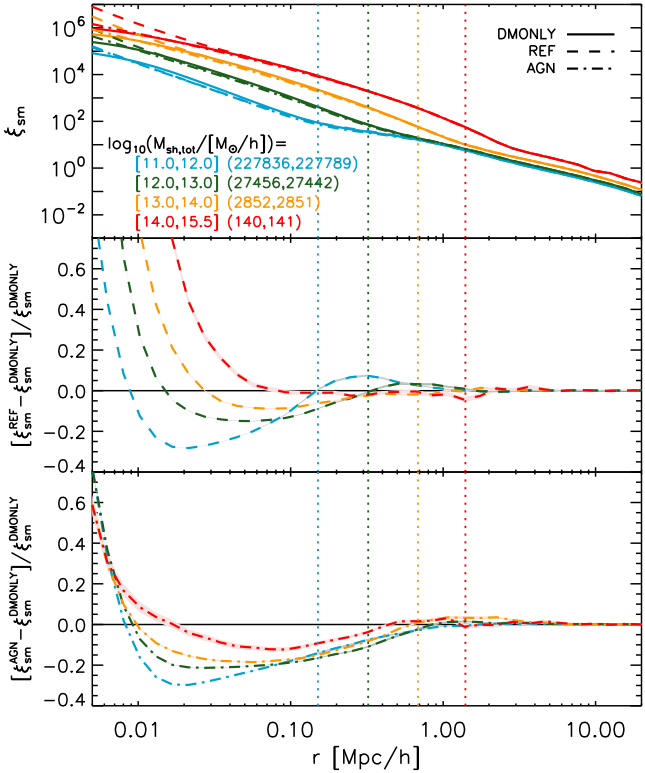}
\caption{As Figure~\ref{sub-mass}, but now only showing the cross-correlation functions between matter and subhaloes that have been linked between a baryonic simulation and \textit{DMONLY}, and that have been selected based on their mass in the latter. Relative to Figure~\ref{sub-mass}, this procedure removes the effects of changes in the subhalo masses, leaving only the effect on the mass profiles and the changes in the positions of the subhaloes. As can be seen from the bottom panel, the change of the mass profile tends to increase the clustering on the very smallest scales (where baryons cool to), but decreases it on intermediate scales (where baryons are evacuated). The latter effect is stronger when AGN feedback is included, and significant over a larger range of scales, for all masses. Shaded areas indicate the regions allowed by 1$\sigma$ bootstrap errors, which are typically much smaller than the widths of the lines.}
\label{sub-mass_linked}
\end{center}
\end{figure}

As we are mainly interested in how galaxy formation changes the clustering of objects with respect to a dark matter only scenario, we use the linking algorithm described in \S\ref{subsec:linking} to link subhaloes in \textit{REF} and \textit{AGN} to those in \textit{DMONLY}, and assign all objects the mass of their \textit{DMONLY} counterpart. Note that this means that there are in fact two different \textit{DMONLY} versions of each correlation function: one derived using all subhaloes for which a counterpart was found in \textit{REF}, and one derived using all subhaloes for which a counterpart was found in \textit{AGN}. In practice, however, the linked halo samples are nearly identical, and the resulting correlation functions for \textit{DMONLY} are virtually indistinguishable. We therefore show only one of these in the top panels of Figures~\ref{sub-sub_linked} and \ref{sub-mass_linked}, although both are used to determine the differences with respect to \textit{REF} and \textit{AGN}.

\subsubsection{Autocorrelation of linked subhaloes}
\label{subsubsec:linkedauto}
We first consider Figure~\ref{sub-sub_linked}, where we show the impact of galaxy formation on the clustering of subhaloes once the change in mass has been accounted for. Comparing first the sample sizes (numbers in the legend) to those in Figure~\ref{sub-sub}, we see that nearly all \textit{DMONLY} subhaloes have a match in each of the baryonic simulations.\footnote{As we now select subhaloes by the mass of their \textit{DMONLY} counterpart, the number of subhaloes can only be directly compared to those of \textit{DMONLY} in Figure~\ref{sub-sub}, not to the number of baryonic subhaloes in Figure~\ref{sub-sub}.} Note that the first number in the legend now indicates the sample size of subhaloes linked between \textit{DMONLY} and \textit{REF}, while the second gives the number of subhaloes linked between \textit{DMONLY} and \textit{AGN}.

We have now also performed 500 bootstrap resamplings for each pair of simulations, and show the 1$\sigma$ errors derived from these as shaded areas in the figure. As we are now using the exact same (linked) sample of subhaloes for any pair of simulations, we are able to avoid overestimating the errors due to the false assumption that the halo samples of the simulations are independent. Similar errors are expected for Figure~\ref{sub-sub}.

Comparing the bottom two panels of Figure~\ref{sub-sub_linked} to those of Figure~\ref{sub-sub}, we immediately see that essentially nothing of the $\sim 10\%$ difference in the clustering amplitude on large scales remains, confirming that this was solely due to galaxy formation changing the masses of these subhaloes. By accounting for the change in the masses of objects due to the effects of baryon physics, one will therefore automatically obtain the correct autocorrelation function at all halo masses, on scales $r \gg r_\mathrm{vir}$.

However, on smaller scales the changes in the dynamics of subhaloes in the baryonic runs become important. This is especially the case for low-mass objects, which are often satellites. As we discussed in \S\ref{subsubsec:subhaloesauto}, \citet{SchewtschenkoMaccio2011} have shown that, initially, satellites in dark matter only simulations move in closer to the centre of the main halo in the same amount of time, which is due in part to the decrease in the virial radius of the main halo when baryons are included (also found for baryonic haloes in our simulations, see \citealp{Velliscig2014}), and in part to the increased dynamical friction experienced by the more massive dark matter satellites. However, as the satellites undergo tidal stripping, baryonic subhaloes are able to retain more of their mass due to their increased concentrations, which causes the situation to reverse on small scales, increasing the number density of baryonic subhaloes relative to pure dark matter ones. This was also found by e.g.\ \citet{Maccio2006}, \citet{Libeskind2010} and \citet{Romano-Diaz2010}. However, at the same time one expects to see an increase in the number density -- and consequently, the clustering -- of baryonic satellite subhaloes at small scales due to the ability to trace baryonic subhaloes longer during infall. This resolution effect could lead to a bias at small separations.

To account for this potential bias, we consider the fraction of subhaloes in \textit{DMONLY} for which a link could be found in \textit{REF} in Appendix~\ref{sec:linkedfractions}. There we show that the fraction of linked subhaloes decreases strongly on small scales for low-mass subhaloes. Higher-resolution simulations are needed to investigate whether the increased survival rate of baryonic subhaloes, and the resulting increase in clustering seen in Figures~\ref{sub-sub} and \ref{sub-sub_linked} on scales $r \la r_\mathrm{vir}$, is physical or not. We have therefore greyed out the curves in these figures on scales where this bias may play a significant role.

However, even after accounting for this potential bias, interesting differences in clustering remain on scales $r \la 2r_\mathrm{vir}$, as Figure~\ref{sub-sub_linked} shows. Especially in the \textit{AGN} simulation, subhaloes tend to be $\sim 10\%$ less clustered at $r \sim r_\mathrm{vir}$. A very small increase in clustering ($\sim 1\%$) can be seen on slightly larger scales, $r \sim 3-4r_\mathrm{vir}$. Both these differences could be explained by the combination of the greater dynamical friction initially experienced by dark matter only subhaloes, together with the delayed infall of baryonic subhaloes. We plan to investigate these effects further in a follow-up paper where we consider the differences in the satellite profiles due to galaxy formation.

Note that small changes in the simulation code (such as changing the level of optimisation when compiling the simulation code) can shift the positions of satellite galaxies and subhaloes by small amounts, even if we start from identical initial conditions.\footnote{The \textit{rms} shift in position for subhaloes between \textit{DMONLY} and \textit{AGN} is about $0.04\,r_\mathrm{vir}$. Similar values are found for shifts between subhaloes in \textit{DMONLY} and \textit{REF}.} However, as almost all these shifts are random, they average out for two-point statistics. Shifts due to dynamical friction and similar effects acting on satellites are the exceptions, as these tend to systematically move satellite subhaloes closer to their respective centrals.

\subsubsection{Cross-correlation with matter}
\label{subsubsec:linkedcross}
Finally, we consider what remains of the baryonic effects on the subhalo-matter cross-correlation function after accounting for the change in the masses of subhaloes. Here, too, we show 1$\sigma$ errors in all panels, now derived from $10000$ bootstrap resamplings. In many cases, the errors are smaller than the widths of the lines.

Comparing the bottom panels of Figure~\ref{sub-mass_linked} to those of Figure~\ref{sub-mass}, we see that while the large-scale offset is now completely removed, we are left with a non-negligible effect on scales $r \la 1\runit$ for all subhalo masses. This again shows the strong effect that feedback can have on the mass distribution: both supernova and AGN feedback move matter to large scales, decreasing $\xi_\mathrm{gm}(r)$. We see that, especially when AGN feedback is included, this can significantly affect clustering out to several times the virial radius, which matches the findings of \citet{vanDaalen2011} and \citet{Velliscig2014}. Note that this also confirms that the findings of \citet{vanDaalen2011}, namely that AGN feedback decreases the matter power spectrum at the $1-10\%$ level out to extremely large scales ($r \sim 10\runit$), are caused by the effect (in Fourier space) of a systematic change in the profile of haloes, rather than by AGN somehow having a significant effect the mass distribution out to more than $10$ times the virial radius of the haloes they occupy.

There are strong similarities between the relative differences that remain for $\xi_\mathrm{sm}$ and the relative differences of halo profiles shown in \citet{Velliscig2014} for the same models, leaving no doubts as to the origin of the signal we see here. The strength of the baryonic effect decreases with increasing mass, but is still highly significant at the mass scales of groups and clusters, although it does not extend beyond the virial radius for the highest-mass subhaloes. The lowest-mass subhaloes we consider here experience a maximum decrease in the cross-correlation with matter of $30\%$, relative to a dark matter only scenario, and even the most massive subhaloes are $10\%$ less strongly clustered with the matter distribution around $r=100\runitk$ when AGN are included. On the smallest scales, the increased clustering due to the cooling of baryons still dominates. Note also that the small-scale differences that we found in Figure~\ref{sub-mass} between \textit{REF} and \textit{AGN} remain.

These results show us that assigning subhaloes in a dark matter only simulation the masses they would have had if galaxy formation and efficient feedback had been included, allows one to obtain the correct clustering predictions on scales $r \gg 1\runit$. However, on smaller scales one cannot correctly predict the cross-correlation with matter, and hence the galaxy-galaxy lensing signal, to better than $\sim 10\%$ accuracy without taking into account the change in the mass distribution.

\section{Summary}
\label{sec:summary}
In this work we investigated how the galaxy and subhalo two-point autocorrelation functions and the cross-correlations with the matter, a measure of the galaxy-galaxy lensing signal, are modified by processes associated with galaxy formation. We utilised a set of cosmological, hydrodynamical simulations with models from the OWLS project, run with more particles and an updated cosmology relative to previous OWLS simulations, to examine what the combined effects on the auto- and cross-correlation functions are of adding baryons and radiative (metal-line) cooling, star formation, chemical enrichment and supernova feedback to a dark matter only simulation, as well as the further addition of a prescription of AGN feedback that reproduces observations of groups and clusters. As nearly all clustering models employed in the literature rely on pure dark matter distributions, either from N-body simulations or halo model type prescriptions, it is important to quantify just how important the effects of baryons and galaxy formation are.

Our findings can be summarised as follows:

\begin{itemize}
\item The stellar masses of galaxies are strongly decreased by (AGN) feedback at fixed subhalo mass, which in turn tends to greatly increase the clustering of galaxies at fixed stellar mass. More importantly for semi-analytical and halo models, the masses of subhaloes are also significantly decreased by the effects of feedback, the result of which is an increase in clustering of $\sim 10\%$ on scales $r \gg 1\runit$, for the full range of subhalo masses considered here ($M_\mathrm{sh,tot}= 10^{11}-10^{15.5}\munit$). This effect is much stronger on smaller scales.

\item Both the change in subhalo mass and the modified subhalo profiles act to change the subhalo-matter cross-correlation function by $\sim 5\%$ on large scales, and significantly more on sub-Mpc scales. The modulation of the signal is strongly non-monotonic and mass-dependent, with both significant increases and decreases in clustering on different scales.
\end{itemize}

We used the identical initial conditions of our simulations to link each baryonic subhalo with its dark matter only counterpart, allowing us to effectively exclude the effect of galaxy formation on the change in the masses of these objects. Nearly all subhaloes are successfully matched in this way.

\begin{itemize}
\item While accounting for the change in mass of subhaloes removes essentially all of the baryonic effects on the autocorrelation of subhaloes on scales $r \gg r_\mathrm{vir}$, deviations $\sim 10\%$ remain on scales $r \la 2r_\mathrm{vir}$, where $r_\mathrm{vir}$ is the virial radius of the subhalo. We argued that these deviations are mainly caused by the differences in the dynamics of satellites, such as the initially greater dynamical friction experienced by the more massive, recently accreted pure dark matter satellites, and the increased concentration of baryonic subhaloes.

\item Finally, on scales $r \la 1\runit$ strong deviations in the subhalo-matter cross-correlation function remain after accounting for the change in the masses of subhaloes. While on galactic scales ($\la 10\,\runitk$) the clustering of subhaloes with matter is always much higher in a baryonic simulation than in the corresponding dark matter only simulation, the inclusion of baryons results in a significant decrease of the cross-correlation for $r \ga 10\runitk$. These effects are stronger for lower-mass subhaloes, reaching up to $30\%$ for subhaloes with masses $10^{11}<M_\mathrm{sh,tot}/[\mathrm{M}_{\sun}/h]<10^{12}$. When AGN feedback is included, $\xi_\mathrm{sm}$ decreases by $\sim 10\%$ relative to a dark matter only simulation for $r \sim 10^2\runitk$, even for subhalo masses $M_\mathrm{sh,tot}>10^{14}\munit$. Mass- and radius-dependent rescalings of halo profiles which extend to several times the virial radius would be needed to account for this effect in dark matter only simulations.
\end{itemize}

We note that while many of our results rely on a model that includes AGN feedback, other feedback processes may have similar effects on clustering. In principle, any other mechanism that is also effective at high masses, sufficiently reducing the stellar masses of massive galaxies, and allows one to reproduce the global properties of groups and clusters, may show similar effects to those shown here for AGN feedback. For example, a model in which a top-heavy IMF is used in high-pressure environments, such as the OWLS model \textit{DBLIMF}, may have the same qualitative effect on clustering \citep[see e.g.][]{vanDaalen2011}.

We stress that while the effects discussed in this paper will certainly need to be modelled in order to achieve the accuracy needed to interpret upcoming cosmological data sets to their full potential, both our knowledge of the relevant physics involved and the currently achievable resolution in cosmological simulations still allow for significant uncertainty in the clustering measures discussed here. The same holds for quantities such as the halo or cluster mass function: much work is yet to be done before we can converge on a realistic prescription of galaxy formation, with uncertainties small enough to match observations in the era of precision cosmology. Although approaches based on dark matter only models, such as semi-analytical modelling or halo occupation distributions, are able to match the observed galaxy mass function, our results imply that their predictions for galaxy-galaxy and galaxy-mass clustering will have errors greater than $10\%$ on sub-Mpc scales, unless the simulation results are modified to correctly account for the effects of baryons on the distributions of mass and satellites.

\section*{Acknowledgements}
The authors thank Raul Angulo, Marcello Cacciato and Simon White for useful comments and discussions, and Marcello Cacciato also for comments on the manuscript. We would also like to thank the anonymous referee for suggestions that improved this paper. The simulations presented here were run on the Cosmology Machine at the Institute for Computational Cosmology in Durham (which is part of the DiRAC Facility jointly funded by STFC, the Large Facilities Capital Fund of BIS, and Durham University) as part of the Virgo Consortium research programme. This work was sponsored by the Dutch National Computing Facilities Foundation (NCF) for the use of supercomputer facilities, with financial support from the Netherlands Organization for Scientific Research (NWO). We also gratefully acknowledge support from the European Research Council under the European Union's Seventh Framework Programme (FP7/2007-2013) / ERC Grant agreement 278594-GasAroundGalaxies and from the Marie Curie Training Network CosmoComp (PITN-GA-2009- 238356). 
\bibliographystyle{mn2e}
\bibliography{PhDbib}

\begin{thebibliography}{69}
\expandafter\ifx\csname natexlab\endcsname\relax\def\natexlab#1{#1}\fi

\bibitem[{{Abadi} {et~al}\mbox{.}(2010){Abadi}, {Navarro}, {Fardal}, {Babul},
  \& {Steinmetz}}]{Abadi2010}
{Abadi} M.~G., {Navarro} J.~F., {Fardal} M., {Babul} A., {Steinmetz} M., 2010,
  \mnras, 407, 435

\bibitem[{{Balaguera-Antol{\'{\i}}nez} \&
  {Porciani}(2013)}]{Balaguera-Antolinez2013}
{Balaguera-Antol{\'{\i}}nez} A., {Porciani} C., 2013, \jcap, 4, 22

\bibitem[{{Balogh} {et~al}\mbox{.}(2001){Balogh}, {Pearce}, {Bower}, \&
  {Kay}}]{Balogh2001}
{Balogh} M.~L., {Pearce} F.~R., {Bower} R.~G., {Kay} S.~T., 2001, \mnras, 326,
  1228

\bibitem[{{Baugh}(2006)}]{Baugh2006}
{Baugh} C.~M., 2006, Reports on Progress in Physics, 69, 3101

\bibitem[{{Behroozi}, {Conroy} \& {Wechsler}(2010){Behroozi}, {Conroy}, \&
  {Wechsler}}]{Behroozi2010}
{Behroozi} P.~S., {Conroy} C., {Wechsler} R.~H., 2010, \apj, 717, 379

\bibitem[{{Berlind} \& {Weinberg}(2002)}]{BerlindWeinberg2002}
{Berlind} A.~A., {Weinberg} D.~H., 2002, \apj, 575, 587

\bibitem[{{Booth} \& {Schaye}(2009)}]{BoothSchaye2009}
{Booth} C.~M., {Schaye} J., 2009, \mnras, 398, 53

\bibitem[{{Bryan} {et~al}\mbox{.}(2013){Bryan}, {Kay}, {Duffy}, {Schaye},
  {Vecchia}, \& {Booth}}]{Bryan2013}
{Bryan} S.~E., {Kay} S.~T., {Duffy} A.~R., {Schaye} J., {Vecchia} C.~D.,
  {Booth} C.~M., 2013, \mnras, 429, 3316

\bibitem[{{Casarini} {et~al}\mbox{.}(2011){Casarini}, {Macci{\`o}},
  {Bonometto}, \& {Stinson}}]{Casarini2011}
{Casarini} L., {Macci{\`o}} A.~V., {Bonometto} S.~A., {Stinson} G.~S., 2011,
  \mnras, 412, 911

\bibitem[{{Conroy}, {Wechsler} \& {Kravtsov}(2006){Conroy}, {Wechsler}, \&
  {Kravtsov}}]{Conroy2006}
{Conroy} C., {Wechsler} R.~H., {Kravtsov} A.~V., 2006, \apj, 647, 201

\bibitem[{{Cooray} \& {Sheth}(2002)}]{CooraySheth2002}
{Cooray} A., {Sheth} R., 2002, \physrep, 372, 1

\bibitem[{{Cui} {et~al}\mbox{.}(2012){Cui}, {Borgani}, {Dolag}, {Murante}, \&
  {Tornatore}}]{Cui2012}
{Cui} W., {Borgani} S., {Dolag} K., {Murante} G., {Tornatore} L., 2012, \mnras,
  423, 2279

\bibitem[{{Dalla Vecchia} \& {Schaye}(2008)}]{DallaVecchiaSchaye2008}
{Dalla Vecchia} C., {Schaye} J., 2008, \mnras, 387, 1431

\bibitem[{{Dolag} {et~al}\mbox{.}(2009){Dolag}, {Borgani}, {Murante}, \&
  {Springel}}]{Dolag2009}
{Dolag} K., {Borgani} S., {Murante} G., {Springel} V., 2009, \mnras, 399, 497

\bibitem[{{Duffy} {et~al}\mbox{.}(2010){Duffy}, {Schaye}, {Kay}, {Dalla
  Vecchia}, {Battye}, \& {Booth}}]{Duffy2010}
{Duffy} A.~R., {Schaye} J., {Kay} S.~T., {Dalla Vecchia} C., {Battye} R.~A.,
  {Booth} C.~M., 2010, \mnras, 405, 2161

\bibitem[{{Gnedin} {et~al}\mbox{.}(2004){Gnedin}, {Kravtsov}, {Klypin}, \&
  {Nagai}}]{Gnedin2004}
{Gnedin} O.~Y., {Kravtsov} A.~V., {Klypin} A.~A., {Nagai} D., 2004, \apj, 616,
  16

\bibitem[{{Governato} {et~al}\mbox{.}(2012){Governato}, {Zolotov}, {Pontzen},
  {Christensen}, {Oh}, {Brooks}, {Quinn}, {Shen}, \& {Wadsley}}]{Governato2012}
{Governato} F. {et~al.}, 2012, \mnras, 422, 1231

\bibitem[{{Guillet}, {Teyssier} \& {Colombi}(2010){Guillet}, {Teyssier}, \&
  {Colombi}}]{Guillet2010}
{Guillet} T., {Teyssier} R., {Colombi} S., 2010, \mnras, 405, 525

\bibitem[{{Guo} {et~al}\mbox{.}(2010){Guo}, {White}, {Li}, \&
  {Boylan-Kolchin}}]{Guo2010}
{Guo} Q., {White} S., {Li} C., {Boylan-Kolchin} M., 2010, \mnras, 404, 1111

\bibitem[{{Haas} {et~al}\mbox{.}(2013){Haas}, {Schaye}, {Booth}, {Dalla
  Vecchia}, {Springel}, {Theuns}, \& {Wiersma}}]{Haas2013}
{Haas} M.~R., {Schaye} J., {Booth} C.~M., {Dalla Vecchia} C., {Springel} V.,
  {Theuns} T., {Wiersma} R.~P.~C., 2013, \mnras, 435, 2931

\bibitem[{{Jing}, {Mo} \& {B\"{o}rner}(1998){Jing}, {Mo}, \&
  {B\"{o}rner}}]{JingMoBoerner1998}
{Jing} Y.~P., {Mo} H.~J., {B\"{o}rner} G., 1998, \apj, 494, 1

\bibitem[{{Jing} {et~al}\mbox{.}(2006){Jing}, {Zhang}, {Lin}, {Gao}, \&
  {Springel}}]{Jing2006}
{Jing} Y.~P., {Zhang} P., {Lin} W.~P., {Gao} L., {Springel} V., 2006, \apjl,
  640, L119

\bibitem[{{Kazantzidis} {et~al}\mbox{.}(2004){Kazantzidis}, {Kravtsov},
  {Zentner}, {Allgood}, {Nagai}, \& {Moore}}]{Kazantzidis2004}
{Kazantzidis} S., {Kravtsov} A.~V., {Zentner} A.~R., {Allgood} B., {Nagai} D.,
  {Moore} B., 2004, \apjl, 611, L73

\bibitem[{{Killedar} {et~al}\mbox{.}(2012){Killedar}, {Borgani}, {Meneghetti},
  {Dolag}, {Fabjan}, \& {Tornatore}}]{Killedar2012}
{Killedar} M., {Borgani} S., {Meneghetti} M., {Dolag} K., {Fabjan} D.,
  {Tornatore} L., 2012, \mnras, 427, 533

\bibitem[{{Komatsu} {et~al}\mbox{.}(2011){Komatsu}, {Smith}, {Dunkley},
  {Bennett}, {Gold}, {Hinshaw}, {Jarosik}, {Larson}, {Nolta}, {Page},
  {Spergel}, {Halpern}, {Hill}, {Kogut}, {Limon}, {Meyer}, {Odegard}, {Tucker},
  {Weiland}, {Wollack}, \& {Wright}}]{Komatsu2011}
{Komatsu} E. {et~al.}, 2011, \apjs, 192, 18

\bibitem[{{Kravtsov} {et~al}\mbox{.}(2004){Kravtsov}, {Berlind}, {Wechsler},
  {Klypin}, {Gottl{\"o}ber}, {Allgood}, \& {Primack}}]{Kravtsov2004}
{Kravtsov} A.~V., {Berlind} A.~A., {Wechsler} R.~H., {Klypin} A.~A.,
  {Gottl{\"o}ber} S., {Allgood} B., {Primack} J.~R., 2004, \apj, 609, 35

\bibitem[{{Landy} \& {Szalay}(1993)}]{LandySzalay1993}
{Landy} S.~D., {Szalay} A.~S., 1993, \apj, 412, 64

\bibitem[{{Libeskind} {et~al}\mbox{.}(2010){Libeskind}, {Yepes}, {Knebe},
  {Gottl{\"o}ber}, {Hoffman}, \& {Knollmann}}]{Libeskind2010}
{Libeskind} N.~I., {Yepes} G., {Knebe} A., {Gottl{\"o}ber} S., {Hoffman} Y.,
  {Knollmann} S.~R., 2010, \mnras, 401, 1889

\bibitem[{{Macci{\`o}} {et~al}\mbox{.}(2006){Macci{\`o}}, {Moore}, {Stadel}, \&
  {Diemand}}]{Maccio2006}
{Macci{\`o}} A.~V., {Moore} B., {Stadel} J., {Diemand} J., 2006, \mnras, 366,
  1529

\bibitem[{{Martizzi} {et~al}\mbox{.}(2013){Martizzi}, {Mohammed}, {Teyssier},
  \& {Moore}}]{Martizzi2013}
{Martizzi} D., {Mohammed} I., {Teyssier} R., {Moore} B., 2013, preprint
  (arXiv:1307.6002)

\bibitem[{{Martizzi} {et~al}\mbox{.}(2012){Martizzi}, {Teyssier}, {Moore}, \&
  {Wentz}}]{Martizzi2012}
{Martizzi} D., {Teyssier} R., {Moore} B., {Wentz} T., 2012, \mnras, 422, 3081

\bibitem[{{McCarthy} {et~al}\mbox{.}(2011){McCarthy}, {Schaye}, {Bower},
  {Ponman}, {Booth}, {Dalla Vecchia}, \& {Springel}}]{McCarthy2011}
{McCarthy} I.~G., {Schaye} J., {Bower} R.~G., {Ponman} T.~J., {Booth} C.~M.,
  {Dalla Vecchia} C., {Springel} V., 2011, \mnras, 412, 1965

\bibitem[{{McCarthy} {et~al}\mbox{.}(2010){McCarthy}, {Schaye}, {Ponman},
  {Bower}, {Booth}, {Dalla Vecchia}, {Crain}, {Springel}, {Theuns}, \&
  {Wiersma}}]{McCarthy2010}
{McCarthy} I.~G. {et~al.}, 2010, \mnras, 406, 822

\bibitem[{{Mead} {et~al}\mbox{.}(2010){Mead}, {King}, {Sijacki}, {Leonard},
  {Puchwein}, \& {McCarthy}}]{Mead2010}
{Mead} J.~M.~G., {King} L.~J., {Sijacki} D., {Leonard} A., {Puchwein} E.,
  {McCarthy} I.~G., 2010, \mnras, 406, 434

\bibitem[{{Moster} {et~al}\mbox{.}(2010){Moster}, {Somerville}, {Maulbetsch},
  {van den Bosch}, {Macci{\`o}}, {Naab}, \& {Oser}}]{Moster2010}
{Moster} B.~P., {Somerville} R.~S., {Maulbetsch} C., {van den Bosch} F.~C.,
  {Macci{\`o}} A.~V., {Naab} T., {Oser} L., 2010, \apj, 710, 903

\bibitem[{{Muldrew}, {Pearce} \& {Power}(2011){Muldrew}, {Pearce}, \&
  {Power}}]{Muldrew2011}
{Muldrew} S.~I., {Pearce} F.~R., {Power} C., 2011, \mnras, 410, 2617

\bibitem[{{Romano-D{\'{\i}}az} {et~al}\mbox{.}(2010){Romano-D{\'{\i}}az},
  {Shlosman}, {Heller}, \& {Hoffman}}]{Romano-Diaz2010}
{Romano-D{\'{\i}}az} E., {Shlosman} I., {Heller} C., {Hoffman} Y., 2010, \apj,
  716, 1095

\bibitem[{{Rudd}, {Zentner} \& {Kravtsov}(2008){Rudd}, {Zentner}, \&
  {Kravtsov}}]{Rudd2008}
{Rudd} D.~H., {Zentner} A.~R., {Kravtsov} A.~V., 2008, \apj, 672, 19

\bibitem[{{Sawala} {et~al}\mbox{.}(2013){Sawala}, {Frenk}, {Crain}, {Jenkins},
  {Schaye}, {Theuns}, \& {Zavala}}]{Sawala2013}
{Sawala} T., {Frenk} C.~S., {Crain} R.~A., {Jenkins} A., {Schaye} J., {Theuns}
  T., {Zavala} J., 2013, \mnras, 431, 1366

\bibitem[{{Scannapieco} {et~al}\mbox{.}(2012){Scannapieco}, {Wadepuhl},
  {Parry}, {Navarro}, {Jenkins}, {Springel}, {Teyssier}, {Carlson}, {Couchman},
  {Crain}, {Dalla Vecchia}, {Frenk}, {Kobayashi}, {Monaco}, {Murante},
  {Okamoto}, {Quinn}, {Schaye}, {Stinson}, {Theuns}, {Wadsley}, {White}, \&
  {Woods}}]{Scannapieco2012}
{Scannapieco} C. {et~al.}, 2012, \mnras, 423, 1726

\bibitem[{{Schaye} \& {Dalla Vecchia}(2008)}]{SchayeDallaVecchia2008}
{Schaye} J., {Dalla Vecchia} C., 2008, \mnras, 383, 1210

\bibitem[{{Schaye} {et~al}\mbox{.}(2010){Schaye}, {Dalla Vecchia}, {Booth},
  {Wiersma}, {Theuns}, {Haas}, {Bertone}, {Duffy}, {McCarthy}, \& {van de
  Voort}}]{Schaye2010}
{Schaye} J. {et~al.}, 2010, \mnras, 402, 1536

\bibitem[{{Schewtschenko} \& {Macci{\`o}}(2011)}]{SchewtschenkoMaccio2011}
{Schewtschenko} J.~A., {Macci{\`o}} A.~V., 2011, \mnras, 413, 878

\bibitem[{{Semboloni}, {Hoekstra} \& {Schaye}(2013){Semboloni}, {Hoekstra}, \&
  {Schaye}}]{Semboloni2013}
{Semboloni} E., {Hoekstra} H., {Schaye} J., 2013, \mnras, 434, 148

\bibitem[{{Semboloni} {et~al}\mbox{.}(2011){Semboloni}, {Hoekstra}, {Schaye},
  {van Daalen}, \& {McCarthy}}]{Semboloni2011}
{Semboloni} E., {Hoekstra} H., {Schaye} J., {van Daalen} M.~P., {McCarthy}
  I.~G., 2011, \mnras, 417, 2020

\bibitem[{{Shankar} {et~al}\mbox{.}(2006){Shankar}, {Lapi}, {Salucci}, {De
  Zotti}, \& {Danese}}]{Shankar2006}
{Shankar} F., {Lapi} A., {Salucci} P., {De Zotti} G., {Danese} L., 2006, \apj,
  643, 14

\bibitem[{{Simha} \& {Cole}(2013)}]{Simha2013}
{Simha} V., {Cole} S., 2013, \mnras, 436, 1142

\bibitem[{{Simha} {et~al}\mbox{.}(2012){Simha}, {Weinberg}, {Dav{\'e}},
  {Fardal}, {Katz}, \& {Oppenheimer}}]{Simha2012}
{Simha} V., {Weinberg} D.~H., {Dav{\'e}} R., {Fardal} M., {Katz} N.,
  {Oppenheimer} B.~D., 2012, \mnras, 423, 3458

\bibitem[{{Springel}(2005)}]{Springel2005c}
{Springel} V., 2005, \mnras, 364, 1105

\bibitem[{{Springel}, {Di Matteo} \& {Hernquist}(2005){Springel}, {Di Matteo},
  \& {Hernquist}}]{Springel2005b}
{Springel} V., {Di Matteo} T., {Hernquist} L., 2005, \mnras, 361, 776

\bibitem[{{Springel} {et~al}\mbox{.}(2001){Springel}, {White}, {Tormen}, \&
  {Kauffmann}}]{Springel2001}
{Springel} V., {White} S.~D.~M., {Tormen} G., {Kauffmann} G., 2001, \mnras,
  328, 726

\bibitem[{{Stanek}, {Rudd} \& {Evrard}(2009){Stanek}, {Rudd}, \&
  {Evrard}}]{Stanek2009}
{Stanek} R., {Rudd} D., {Evrard} A.~E., 2009, \mnras, 394, L11

\bibitem[{{Stott} {et~al}\mbox{.}(2012){Stott}, {Hickox}, {Edge}, {Collins},
  {Hilton}, {Harrison}, {Romer}, {Rooney}, {Kay}, {Miller}, {Sahl{\'e}n},
  {Lloyd-Davies}, {Mehrtens}, {Hoyle}, {Liddle}, {Viana}, {McCarthy}, {Schaye},
  \& {Booth}}]{Stott2012}
{Stott} J.~P. {et~al.}, 2012, \mnras, 422, 2213

\bibitem[{{Tinker} {et~al}\mbox{.}(2005){Tinker}, {Weinberg}, {Zheng}, \&
  {Zehavi}}]{Tinker2005}
{Tinker} J.~L., {Weinberg} D.~H., {Zheng} Z., {Zehavi} I., 2005, \apj, 631, 41

\bibitem[{{Tissera} {et~al}\mbox{.}(2010){Tissera}, {White}, {Pedrosa}, \&
  {Scannapieco}}]{Tissera2010}
{Tissera} P.~B., {White} S.~D.~M., {Pedrosa} S., {Scannapieco} C., 2010,
  \mnras, 406, 922

\bibitem[{{Vale} \& {Ostriker}(2004)}]{ValeOstriker2004}
{Vale} A., {Ostriker} J.~P., 2004, \mnras, 353, 189

\bibitem[{{van Daalen} {et~al}\mbox{.}(2011){van Daalen}, {Schaye}, {Booth}, \&
  {Dalla Vecchia}}]{vanDaalen2011}
{van Daalen} M.~P., {Schaye} J., {Booth} C.~M., {Dalla Vecchia} C., 2011,
  \mnras, 415, 3649

\bibitem[{{van den Bosch} {et~al}\mbox{.}(2013){van den Bosch}, {More},
  {Cacciato}, {Mo}, \& {Yang}}]{vandenBosch2013}
{van den Bosch} F.~C., {More} S., {Cacciato} M., {Mo} H., {Yang} X., 2013,
  \mnras, 430, 725

\bibitem[{{Velliscig} {et~al}\mbox{.}(2014){Velliscig}, {van Daalen}, {Schaye},
  {McCarthy}, {Cacciato}, {Le Brun}, \& {Dalla Vecchia}}]{Velliscig2014}
{Velliscig} M., {van Daalen} M.~P., {Schaye} J., {McCarthy} I.~G., {Cacciato}
  M., {Le Brun} A.~M.~C., {Dalla Vecchia} C., 2014, preprint (arXiv:1402.4461)

\bibitem[{{Vikhlinin} {et~al}\mbox{.}(2006){Vikhlinin}, {Kravtsov}, {Forman},
  {Jones}, {Markevitch}, {Murray}, \& {Van Speybroeck}}]{Vikhlinin2006}
{Vikhlinin} A., {Kravtsov} A., {Forman} W., {Jones} C., {Markevitch} M.,
  {Murray} S.~S., {Van Speybroeck} L., 2006, \apj, 640, 691

\bibitem[{{Wechsler} {et~al}\mbox{.}(2006){Wechsler}, {Zentner}, {Bullock},
  {Kravtsov}, \& {Allgood}}]{Wechsler2006}
{Wechsler} R.~H., {Zentner} A.~R., {Bullock} J.~S., {Kravtsov} A.~V., {Allgood}
  B., 2006, \apj, 652, 71

\bibitem[{{Weinberg} {et~al}\mbox{.}(2008){Weinberg}, {Colombi}, {Dav{\'e}}, \&
  {Katz}}]{Weinberg2008}
{Weinberg} D.~H., {Colombi} S., {Dav{\'e}} R., {Katz} N., 2008, \apj, 678, 6

\bibitem[{{White}(2004)}]{White2004}
{White} M., 2004, Astroparticle Physics, 22, 211

\bibitem[{{Wiersma}, {Schaye} \& {Smith}(2009){Wiersma}, {Schaye}, \&
  {Smith}}]{Wiersma2009a}
{Wiersma} R.~P.~C., {Schaye} J., {Smith} B.~D., 2009, \mnras, 393, 99

\bibitem[{{Wiersma} {et~al}\mbox{.}(2009){Wiersma}, {Schaye}, {Theuns}, {Dalla
  Vecchia}, \& {Tornatore}}]{Wiersma2009b}
{Wiersma} R.~P.~C., {Schaye} J., {Theuns} T., {Dalla Vecchia} C., {Tornatore}
  L., 2009, \mnras, 399, 574

\bibitem[{{Yang} {et~al}\mbox{.}(2013){Yang}, {Kratochvil}, {Huffenberger},
  {Haiman}, \& {May}}]{Yang2013}
{Yang} X., {Kratochvil} J.~M., {Huffenberger} K., {Haiman} Z., {May} M., 2013,
  \prd, 87, 023511

\bibitem[{{Yang}, {Mo} \& {van den Bosch}(2003){Yang}, {Mo}, \& {van den
  Bosch}}]{YangMovandenBosch2003}
{Yang} X., {Mo} H.~J., {van den Bosch} F.~C., 2003, \mnras, 339, 1057

\bibitem[{{Zentner} {et~al}\mbox{.}(2013){Zentner}, {Semboloni}, {Dodelson},
  {Eifler}, {Krause}, \& {Hearin}}]{Zentner2013}
{Zentner} A.~R., {Semboloni} E., {Dodelson} S., {Eifler} T., {Krause} E.,
  {Hearin} A.~P., 2013, \prd, 87, 043509

\bibitem[{{Zhan} \& {Knox}(2004)}]{ZhanKnox2004}
{Zhan} H., {Knox} L., 2004, \apjl, 616, L75

\end{thebibliography}

\appendix
\section{Convergence tests}
\label{sec:convergence}
\begin{figure*}
\begin{center}
\begin{tabular}{cc}
\includegraphics[width=1.0\columnwidth, trim=12mm 8mm 10mm 7mm]{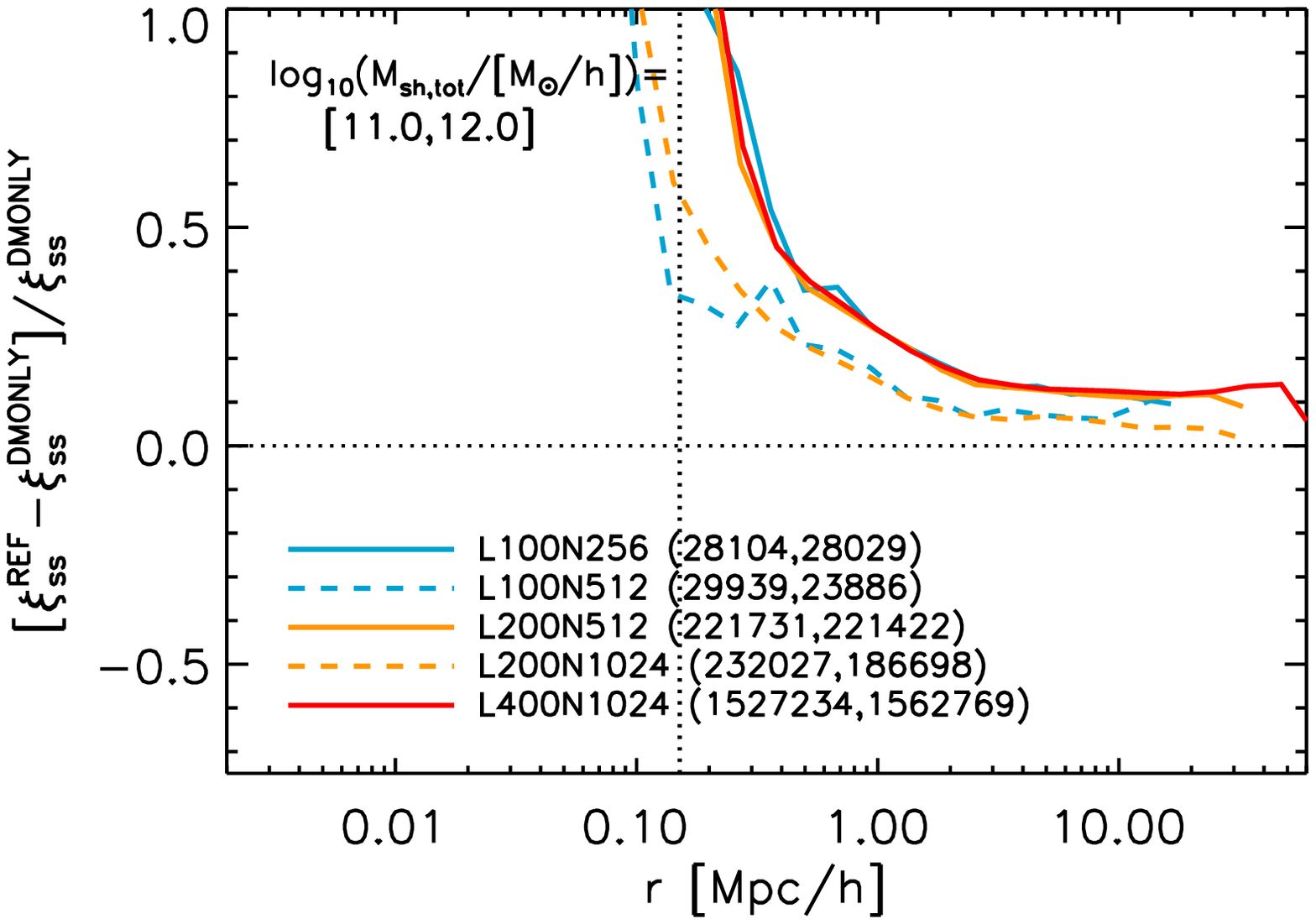} & \includegraphics[width=1.0\columnwidth, trim=12mm 8mm 10mm 7mm]{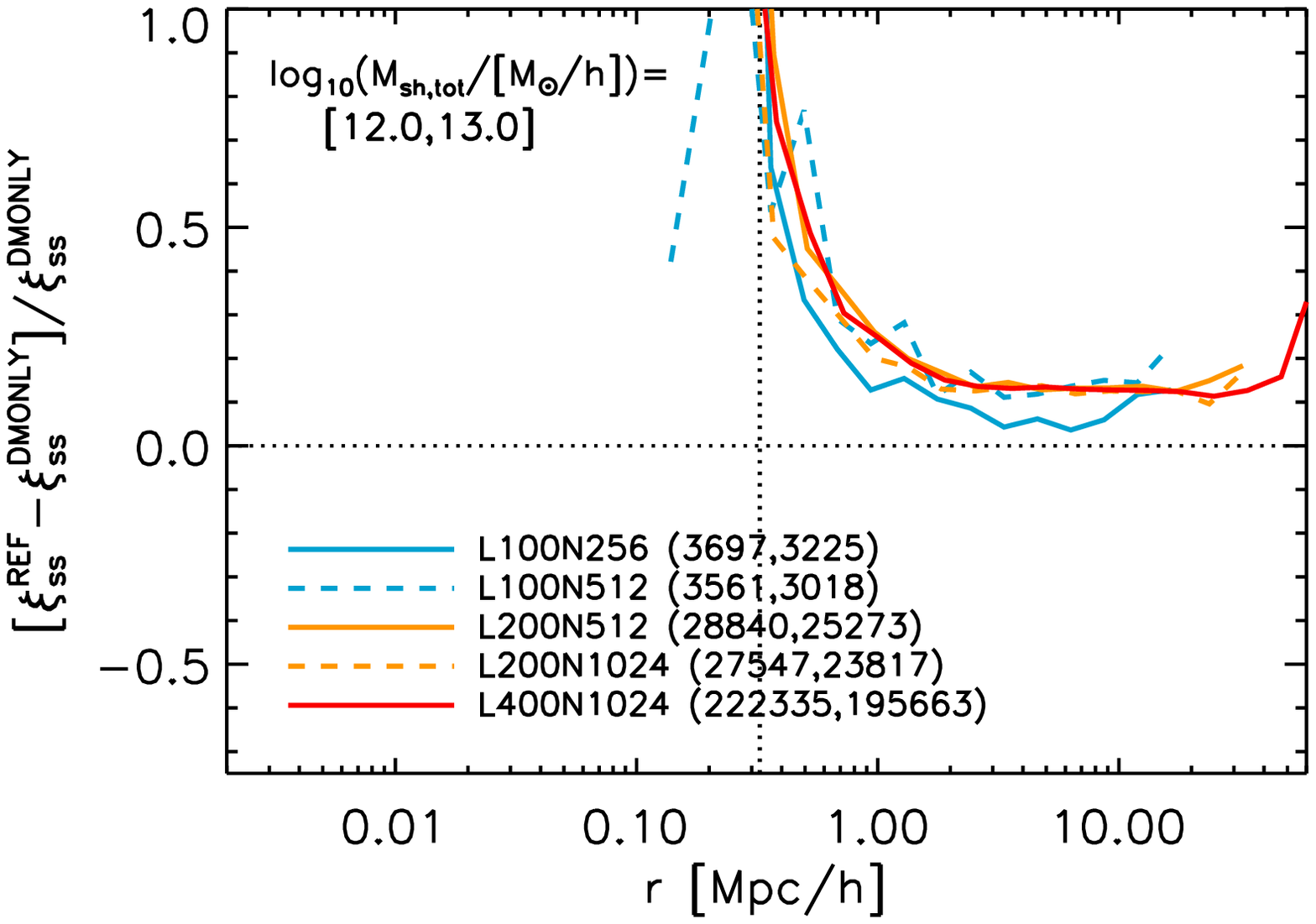}\\
\includegraphics[width=1.0\columnwidth, trim=12mm 8mm 10mm 7mm]{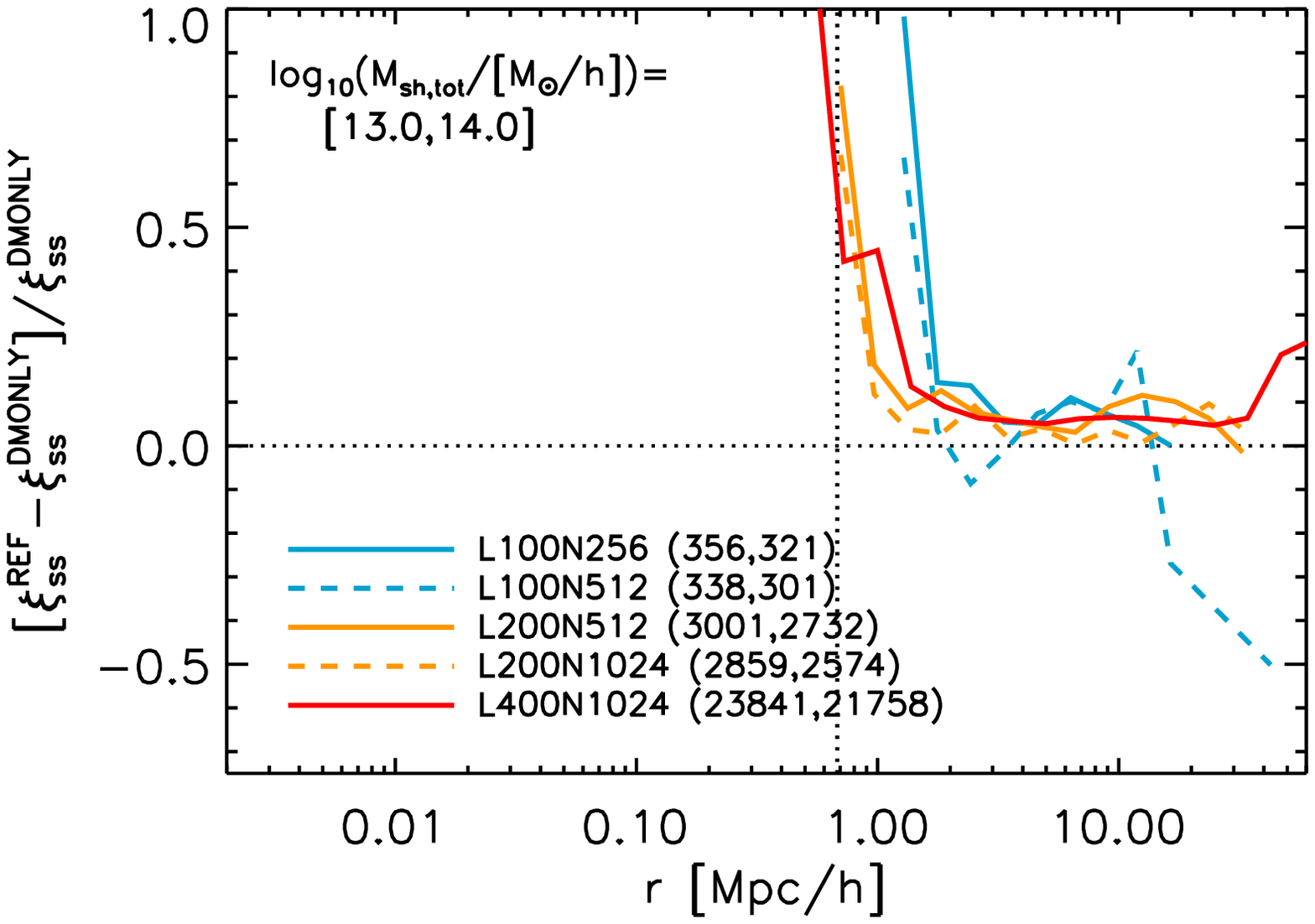} & \includegraphics[width=1.0\columnwidth, trim=12mm 8mm 10mm 7mm]{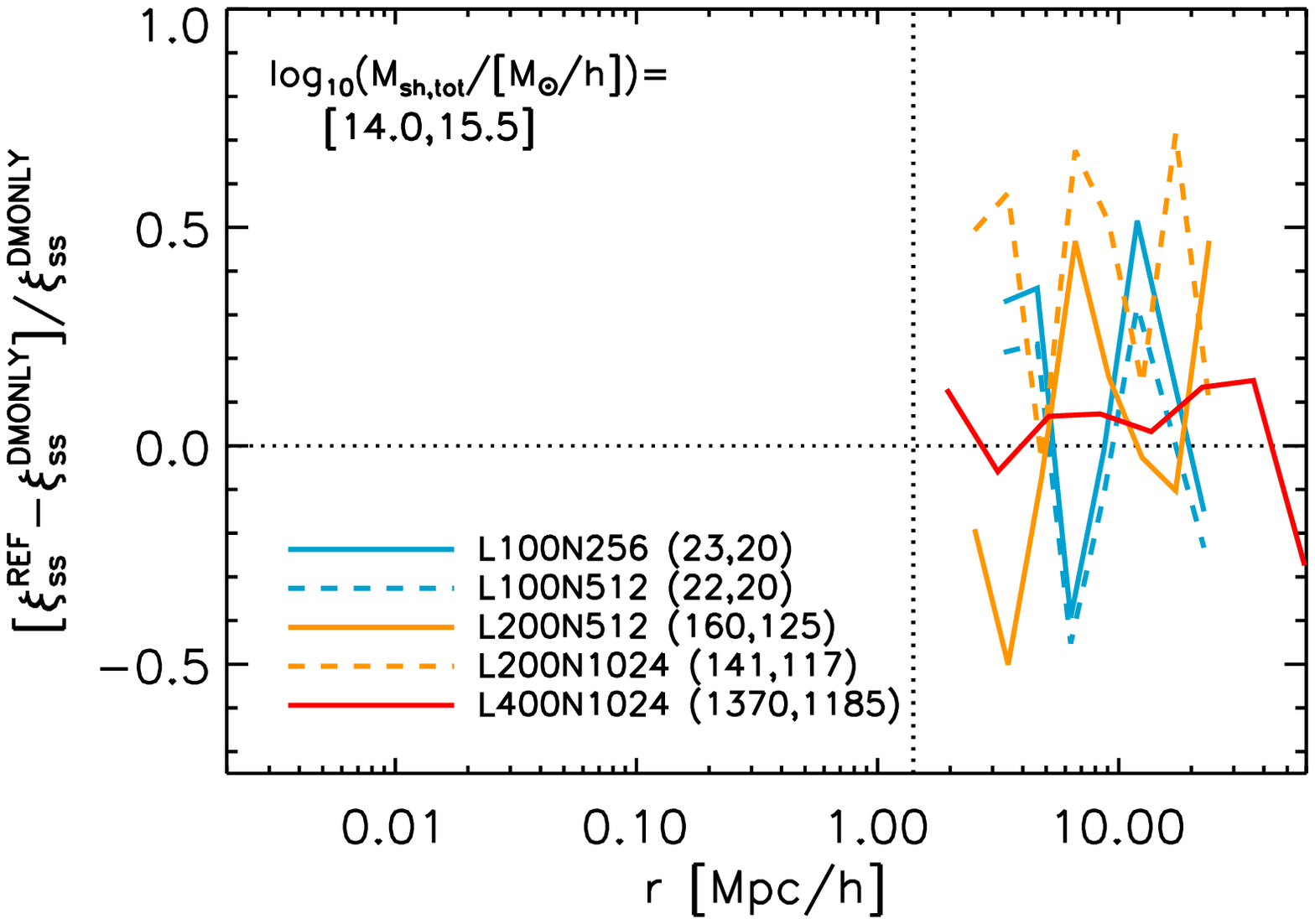}
\end{tabular}
\caption{The relative differences in the subhalo autocorrelation functions between models \textit{DMONLY} and \textit{REF}, split by subhalo mass as indicated in the top left of each panel. Contrary to the plots shown in other sections, no minimum number of pairs per bin is imposed. The box sizes and particle numbers, as well as the subhalo numbers for \textit{DMONLY} and \textit{REF}, respectively, are indicated in the legend, and a vertical dotted line indicates the mean virial radius in each mass bin. At fixed resolution (same line style) very little changes, although the effect of the better statistics offered by a larger volume are apparent. At fixed box size (same colour) the results are also very similar, except for the lowest mass bin, where the small-scale clustering is resolution dependent. Note that all simulations show excellent agreement for $10^{12}<M_\mathrm{sh,tot}/[\mathrm{M}_{\sun}/h]<10^{13}$, where neither resolution nor volume is an issue.}
\label{autorestests}
\end{center}
\end{figure*}
Here we investigate the effects of changing the box size or resolution of the simulations used in this paper on the subhalo autocorrelation function, as this is the main focus of this paper. We will also briefly discuss the effects on the subhalo-matter cross-correlation function.

\begin{figure*}
\begin{center}
\begin{tabular}{cc}
\includegraphics[width=1.0\columnwidth, trim=12mm 8mm 10mm 7mm]{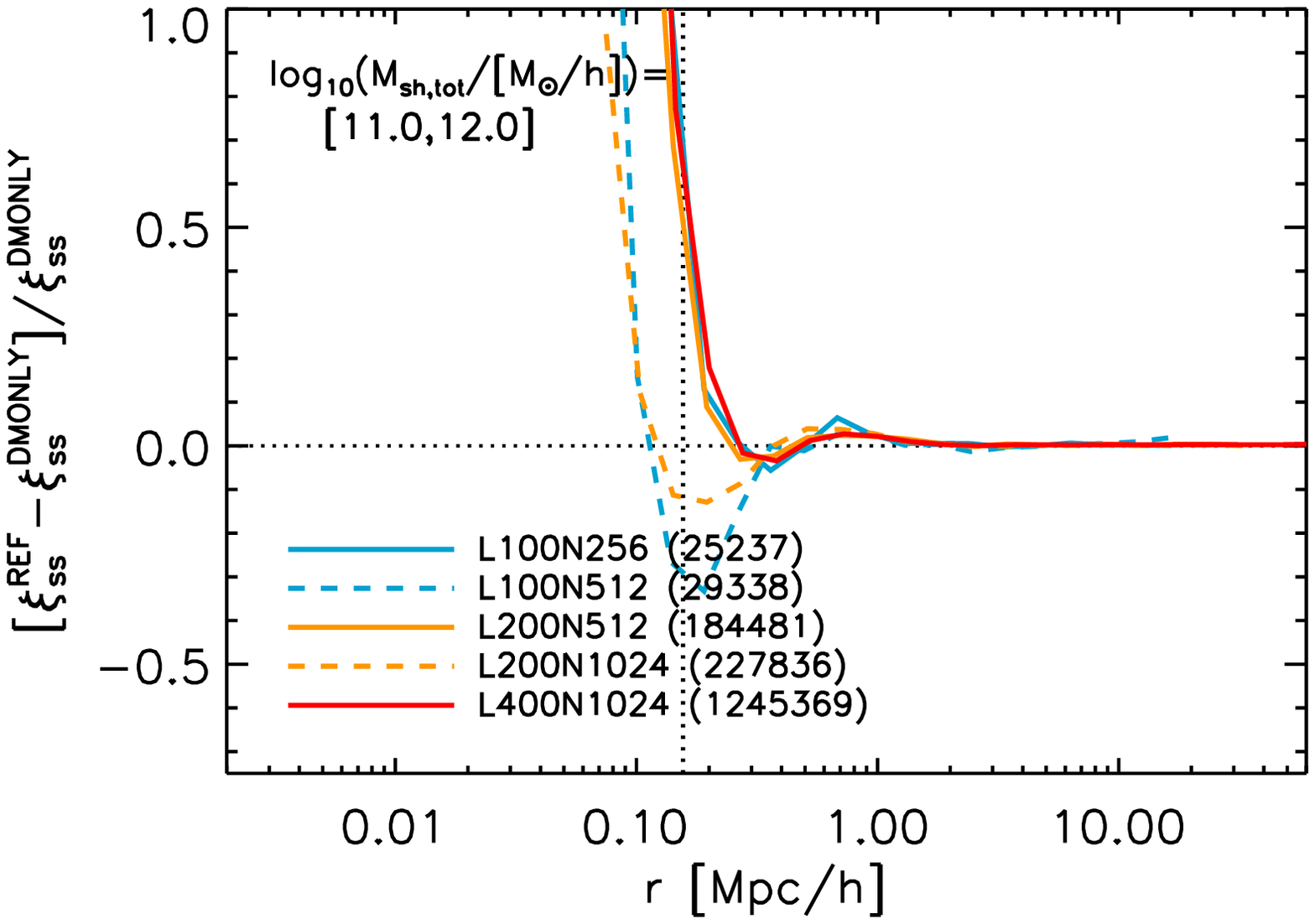} & \includegraphics[width=1.0\columnwidth, trim=12mm 8mm 10mm 7mm]{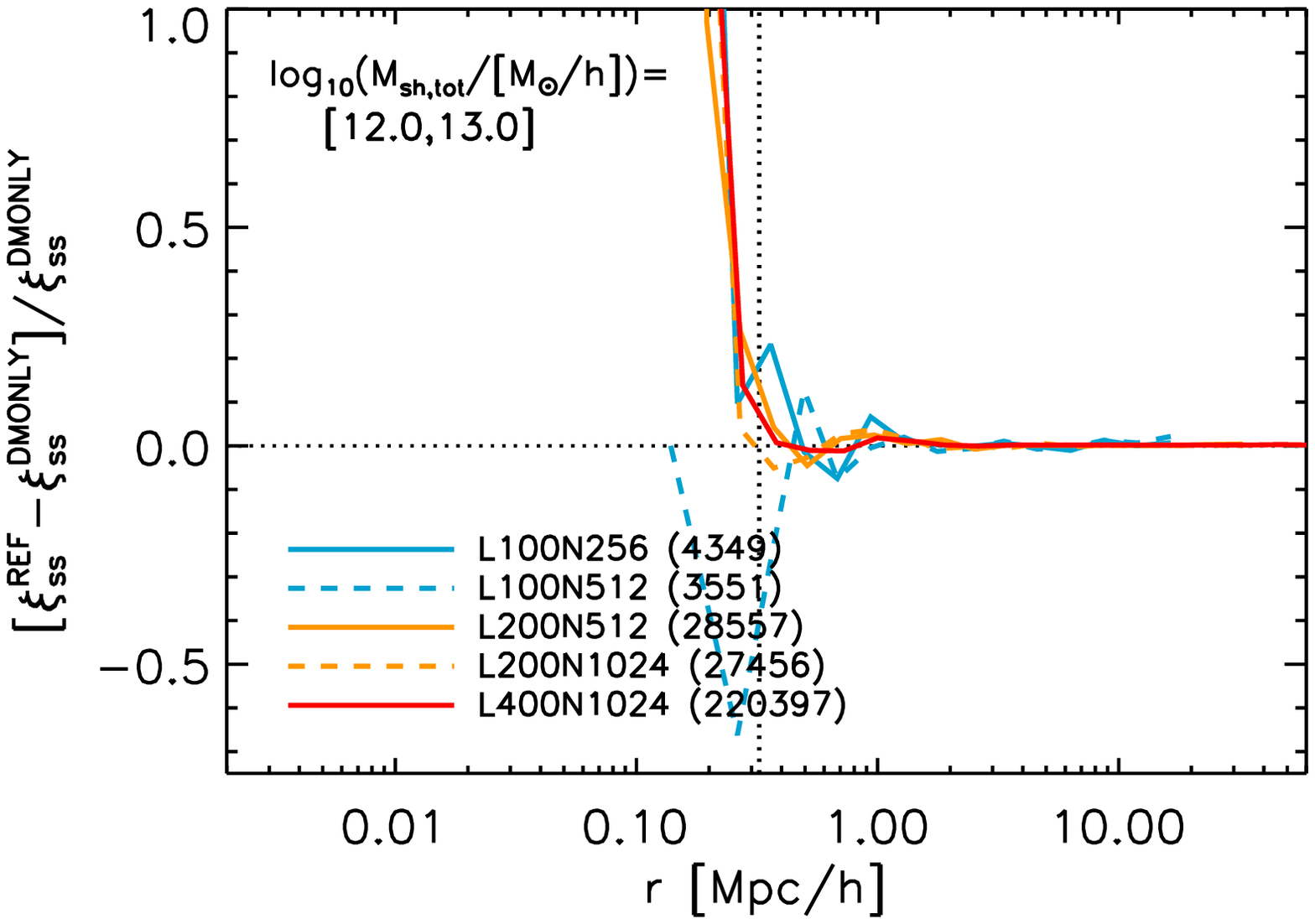}\\
\includegraphics[width=1.0\columnwidth, trim=12mm 8mm 10mm 7mm]{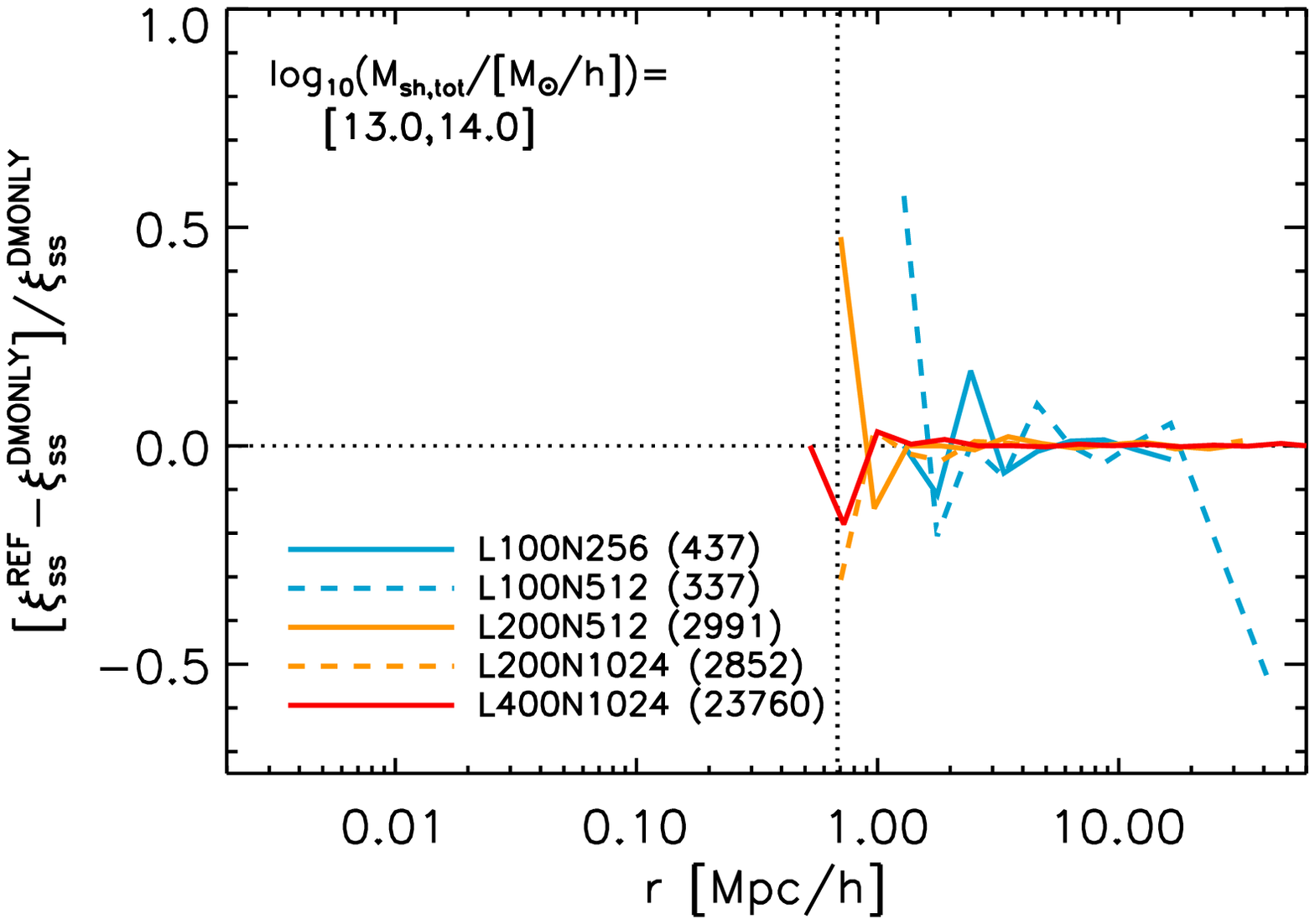} & \includegraphics[width=1.0\columnwidth, trim=12mm 8mm 10mm 7mm]{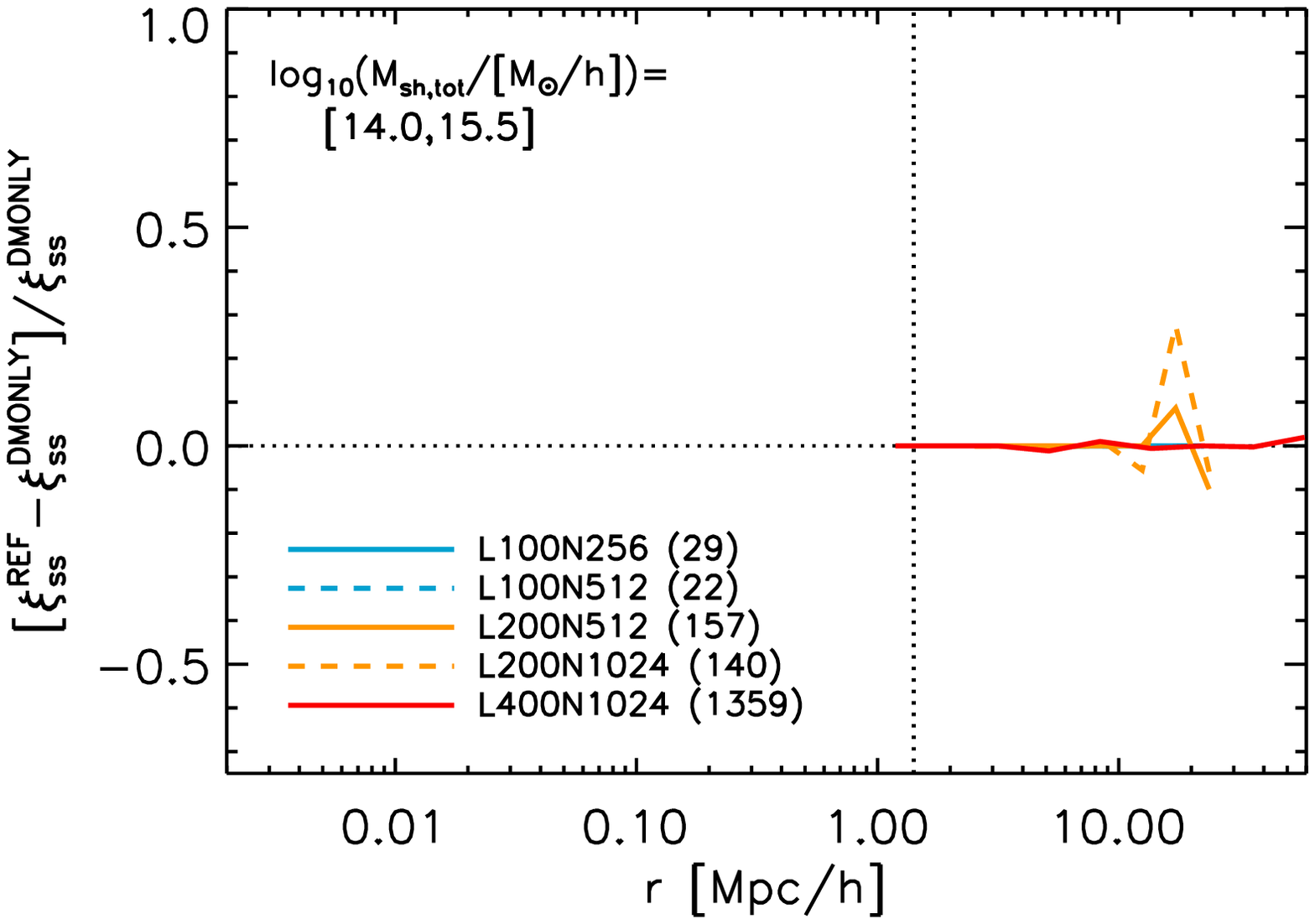}
\end{tabular}
\caption{As Figure~\ref{autorestests}, but now only showing the autocorrelation functions for subhaloes linked between \textit{REF} and \textit{DMONLY}, and selected based on their mass in the latter. The convergence here is very similar to that seen in Figure~\ref{autorestests}.}
\label{autorestests_linked}
\end{center}
\end{figure*}

In Figure~\ref{autorestests} we show the subhalo autocorrelation functions for models \textit{DMONLY} and \textit{REF}. For clarity the correlation functions for the \textit{AGN} model are not shown, but the results are very similar. Contrary to what was done for the figures in the main text, here we do not impose a minimum number of pairs per bin. We vary both the box size and particle number in a systematic way: for simulations shown with the same line style (either solid or dashed) we vary the box size at fixed resolution, while for simulations shown with the same colour we vary the resolution at fixed box size.

We first consider the effect of changing the size of the simulated volume. Looking at the solid and dashed lines separately, we can see that very little changes at fixed resolution, except that the results clearly benefit from the better statistics offered by a larger volume. This is noticeable both for the rare high-mass objects, on any scale, and for low-mass objects on the very smallest scales, where very few pairs are found.

If we instead consider each colour of Figure~\ref{autorestests} separately, we see that at fixed box size the results are also very similar. The exception is the lowest mass bin, $10^{11}<M_\mathrm{sh,tot}/[\mathrm{M}_{\sun}/h]<10^{12}$, where the correlation function is clearly resolution dependent when baryons are included. This is because these subhaloes contain only $\sim 10^2$ particles in the low-resolution simulations, which is not quite enough for convergence, especially when feedback processes are included. We have verified that the subhalo mass functions of the highest-resolution simulations shown here are indeed converged using simulations with smaller volumes and higher resolutions (not shown here). The results for the second mass bin on the other hand, $10^{12}<M_\mathrm{sh,tot}/[\mathrm{M}_{\sun}/h]<10^{13}$, are fully consistent between the different resolutions shown here.

We have repeated these same resolution tests for the autocorrelation functions of linked subhaloes, shown in Figure~\ref{autorestests_linked}. Here, too, we see that our results are converged for $M_\mathrm{sh,tot}>10^{12}\munit$.

Based on these tests, we choose to use the higher-resolution \textit{L200N1024} simulations for subhaloes with masses $10^{11}<M_\mathrm{sh,tot}/[\mathrm{M}_{\sun}/h]<10^{13}$, and take advantage of the better statistics offered by the \textit{L400N1024} simulations for subhalo masses $M_\mathrm{sh,tot}>10^{13}\munit$. Similarly, we opt to use the higher-resolution simulation for the autocorrelation function of galaxies with stellar masses $10^{9}<M_*/[\mathrm{M}_{\sun}/h]<10^{11}$, and the larger-volume simulation for galaxies with $M_*>10^{11}\munit$.

We also verified that the cross-correlation functions shown in this work are sufficiently converged (not shown). For the subhalo-matter (and galaxy-matter) cross correlation functions, statistics are less of an issue, as the number of particles is the same for the \textit{L200} and \textit{L400} simulations. In other words, while for the autocorrelation functions the number of pairs scales as $N_\mathrm{obj}^2$, the number of pairs for the cross-correlation functions scales as $N_\mathrm{obj}N_\mathrm{part}$, where $N_\mathrm{part} \gg N_\mathrm{obj}$. Resolution is still an issue, however: while simulations including baryons always show stronger clustering on galaxy scales than \textit{DMONLY}, the exact scale on which the transition of a relative increase to a relative decrease in clustering occurs depends somewhat on the softening length. Additionally, as we discussed briefly in \S\ref{subsec:galaxies}, the effect of AGN feedback is resolution-dependent in our simulations, due to the fact that seed black holes can only be inserted in resolved haloes. AGN feedback may therefore be weaker at the \textit{L400} resolution than at the \textit{L200} resolution, while the strength of the feedback in the latter was deemed realistic. We therefore choose to use the \textit{L200} simulations \emph{at all masses} when considering the cross-correlation functions $\xi_\mathrm{gm}$ and $\xi_\mathrm{sm}$, valuing resolution over volume.

\section{Linked fractions}
\label{sec:linkedfractions}
\begin{figure*}
\begin{center}
\begin{tabular}{cc}
\includegraphics[width=1.0\columnwidth, trim=12mm 8mm 10mm 7mm]{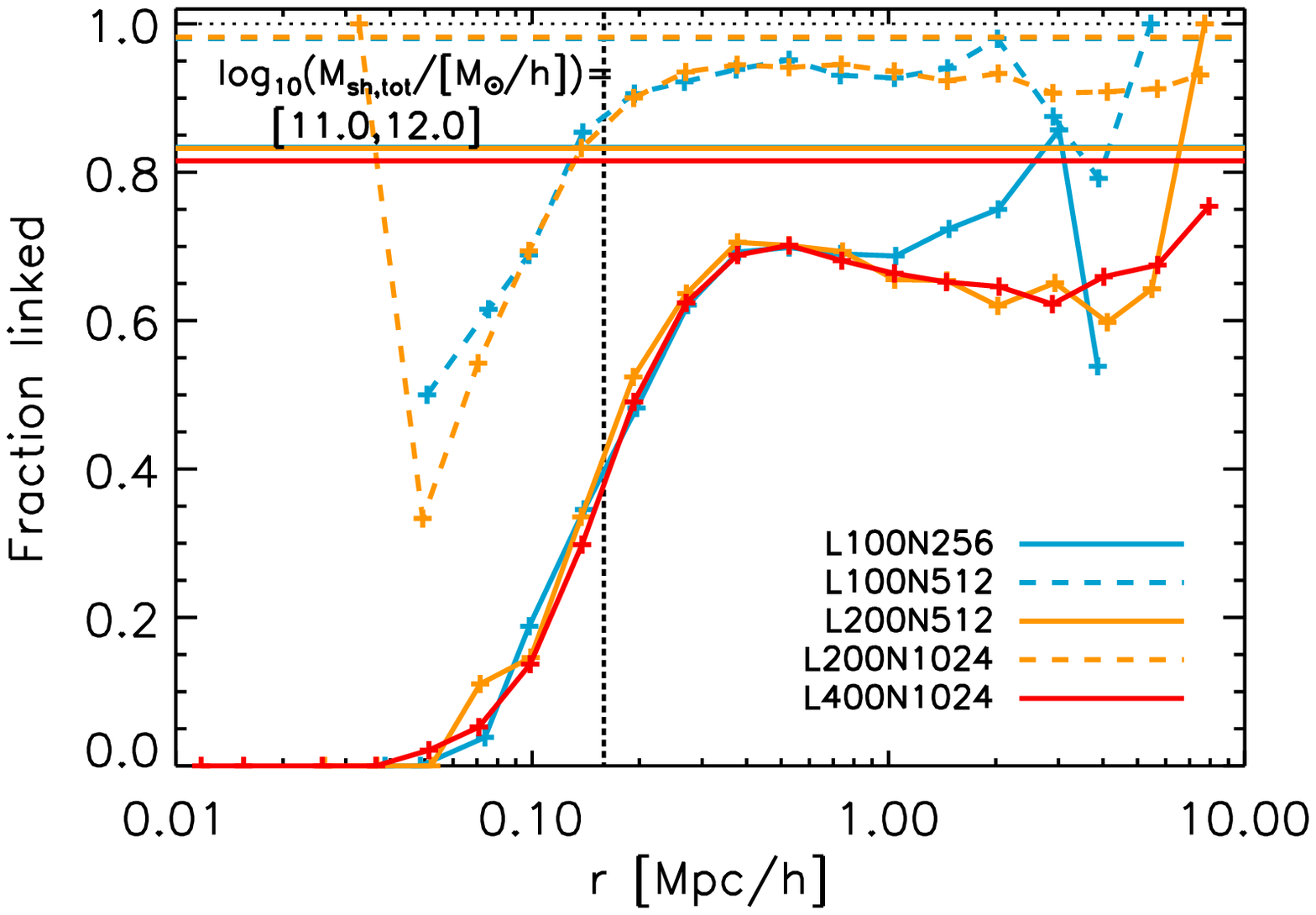} & \includegraphics[width=1.0\columnwidth, trim=12mm 8mm 10mm 7mm]{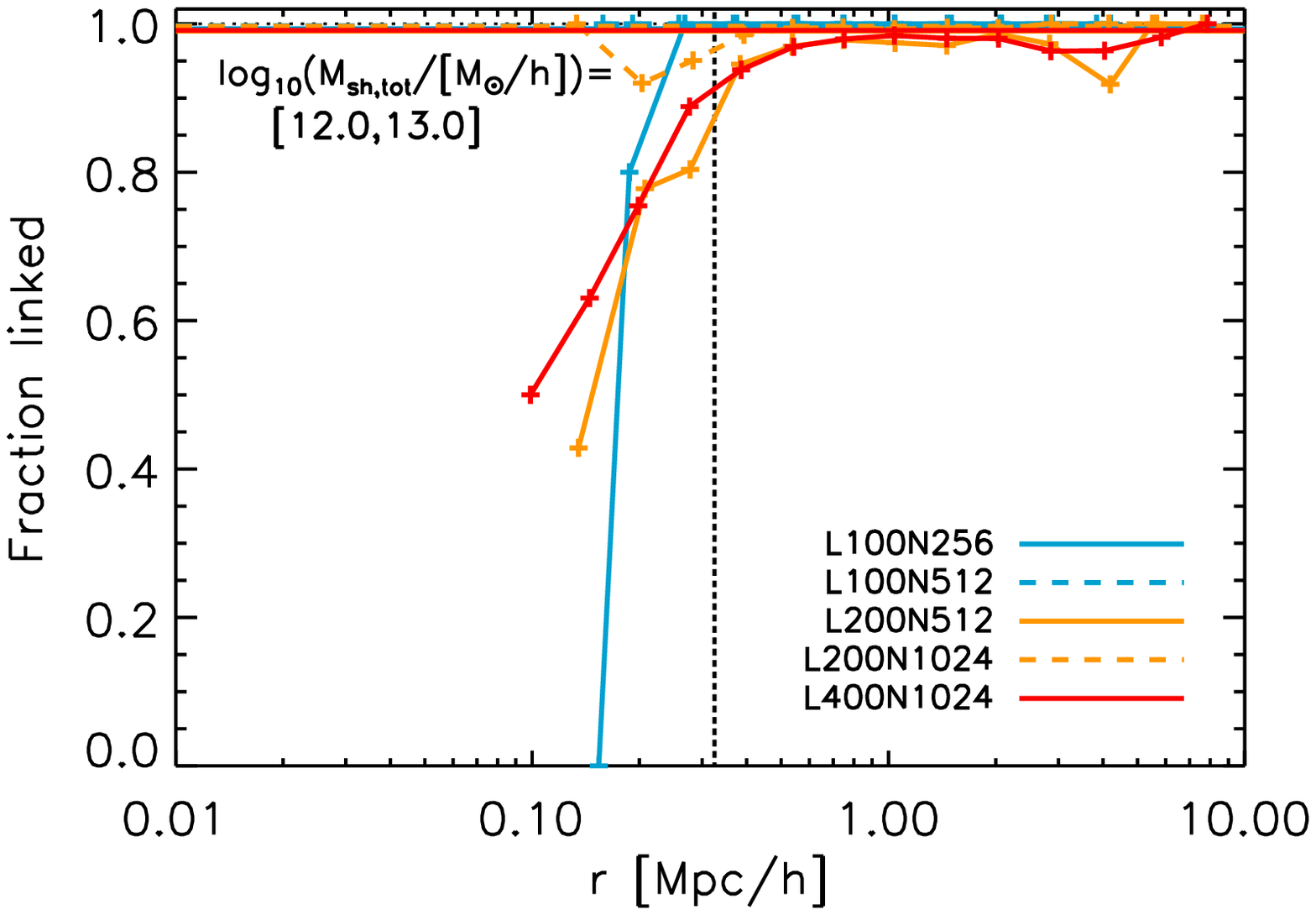}\\
\includegraphics[width=1.0\columnwidth, trim=12mm 8mm 10mm 7mm]{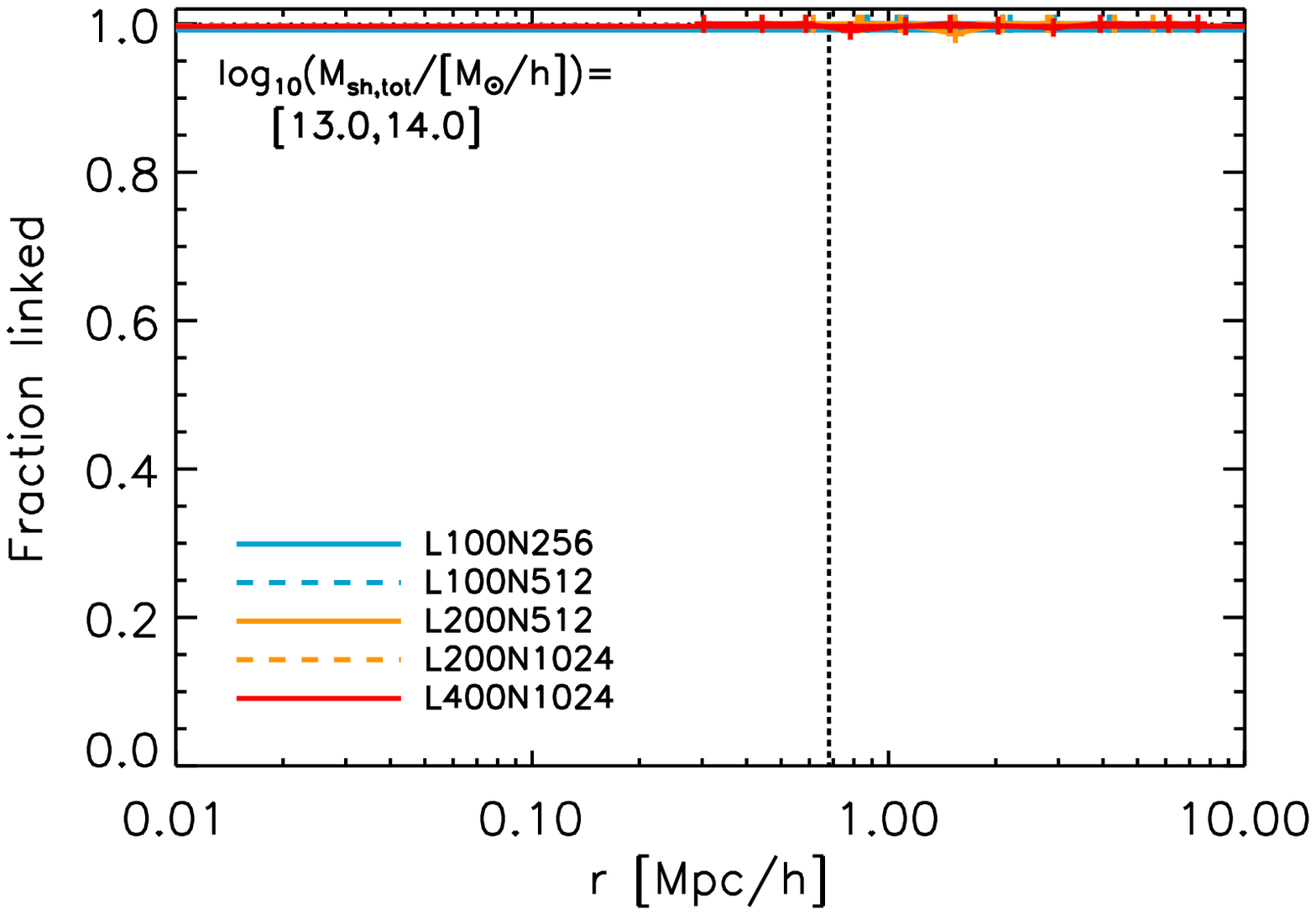} & \includegraphics[width=1.0\columnwidth, trim=12mm 8mm 10mm 7mm]{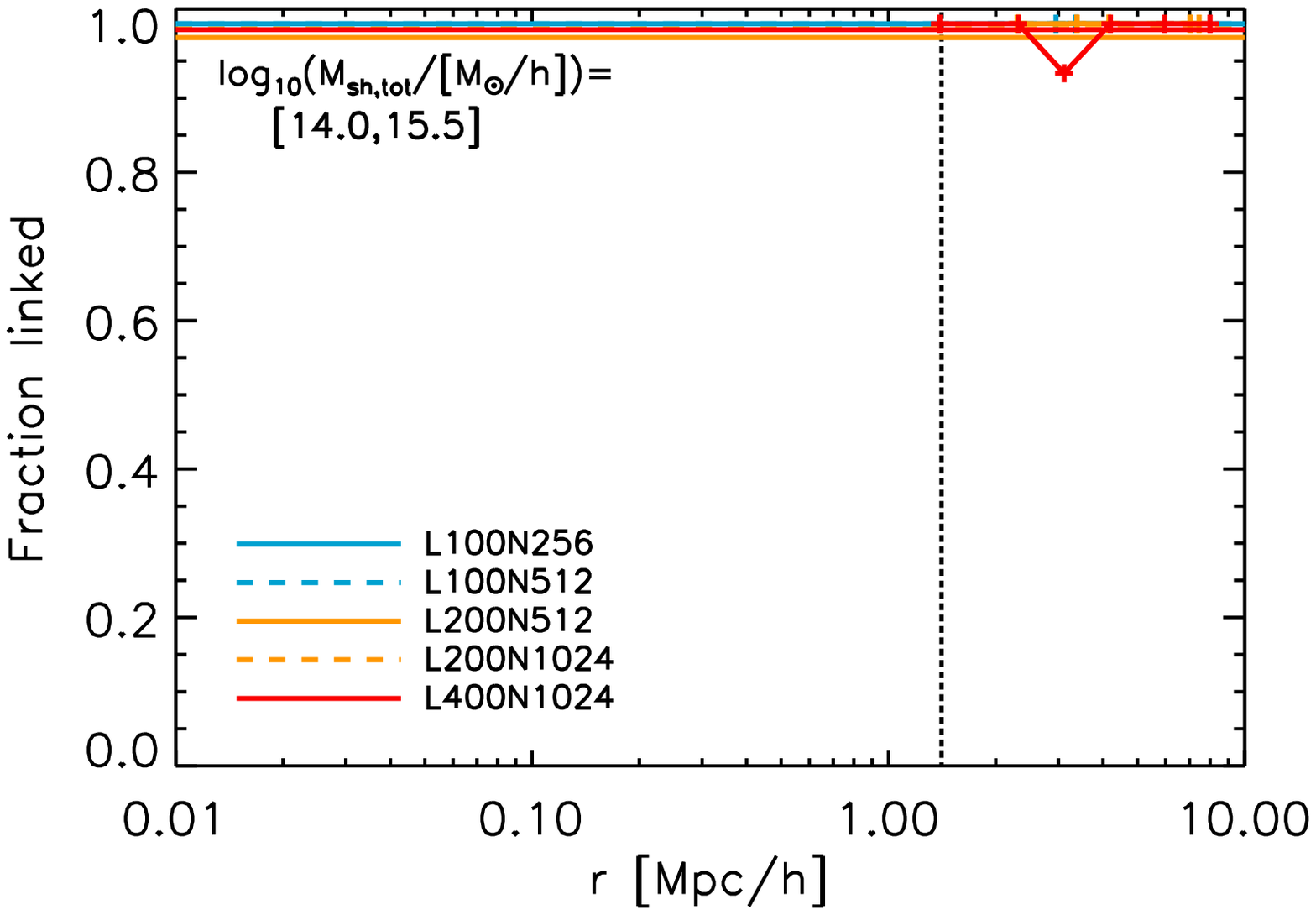}
\end{tabular}
\caption{The fraction of subhaloes in \textit{DMONLY} for which a link was found in \textit{REF}, split by subhalo mass as indicated in the top left of each panel. Colours and line styles are as in Figure~\ref{autorestests}, and a vertical dotted line once again indicates the mean virial radius in each mass bin. Horizontal lines show the total fraction of linked subhaloes (both centrals and satellites) at the corresponding box size and resolution, while the lines with plot symbols show the fraction of satellite subhaloes linked as a function of radius. For $r \la 2r_\mathrm{vir}$ the fraction of linked satellites typically drops sharply as subhaloes are destroyed by tidal stripping or become undetectable. Both the matched satellite and total fractions depends strongly on box size and resolution for subhalo masses $M_\mathrm{sh,tot}<10^{12}\munit$.}
\label{linkedfraction}
\end{center}
\end{figure*}

Here we consider the fraction of subhaloes for which a link can be established between \textit{DMONLY} and \textit{REF} as a function of both mass and, in the case of satellites, radius. Both numerical and physical effects play a role here. First, at small radii \textsc{subfind} may fail to detect satellite subhaloes even though these have not been fully disrupted yet, due to the high background density of the main halo \citep[e.g.][]{Muldrew2011}. As baryonic subhaloes are typically more concentrated than dark matter only ones, increasing their density contrast, these can be detected down to smaller radii. Second, baryonic satellites tend to be survive longer than their dark matter only counterparts, as their increased concentration also allows them to better withstand the tidal forces of the main halo \citep[e.g.][]{Maccio2006}. Because of this, our results for linked samples may be biased at radii where a significant fraction of satellite subhaloes is unlinked, as we expect to be better able to detect a pair of identical subhaloes when the baryonic one is located at smaller radii than the dark matter only one, relative to a situation in which the dark matter only satellite is located at smaller radii than its baryonic counterpart.

In Figure~\ref{linkedfraction} we show the fraction of subhaloes in \textit{DMONLY} for which a counterpart is found in \textit{REF}. Once again we do not show a comparison with \textit{AGN} for clarity, but note that very similar results are obtained.

Horizontal lines show the total fraction of \textit{DMONLY} subhaloes (both centrals and satellites) that is recovered in \textit{REF}, while lines with plot symbols show the fraction of satellites for which a link is found as a function of radius. It is clear that the linked fraction depends heavily on both box size and resolution for $M_\mathrm{sh,tot}<10^{12}\munit$, although the effect of the box size is only significant for the low-resolution simulations. For the simulation employed in this mass bin throughout the main text of the paper, \textit{L200N1024}, the total fraction of linked subhaloes is around $98\%$. However, the fraction of linked satellites is significantly lower, especially for radii $r \la 2r_\mathrm{vir}$, where the different survival and detection rates of baryonic subhaloes are expected to play a role.

Comparing this panel to the corresponding panel in Figure~\ref{autorestests_linked}, we see that the drop in the fraction of matched satellites at small radii corresponds to the strong increase in clustering found for baryonic subhaloes, indicating that this may be a biased result. Similar results are found for satellites with masses $10^{12}<M_\mathrm{sh,tot}/[\mathrm{M}_{\sun}/h]<10^{13}$, although both the total and the satellite linked fractions are much higher than for $10^{12}<M_\mathrm{sh,tot}/[\mathrm{M}_{\sun}/h]<10^{13}$, for all simulations and radii. No drop-off in the linked fraction of satellites is observed at higher masses.

Based on these results, we haven chosen to grey out the relative difference curves in the Figures showing autocorrelation functions (Figures~\ref{sub-sub} and \ref{sub-sub_linked}) on radii where the fraction of linked satellites is $<95\%$ of the total matched fraction. Note that this may not completely remove the possible bias on scales where the satellite contribution dominates the correlation function. Further investigation with higher-resolution simulations is needed to determine whether the upturn observed at small radii is physical or numerical in origin.

Note that the occasional downturn of the linked fraction at relatively large radii, $r \ga 2r_\mathrm{vir}$, is due to small-number statistics, as low-mass subhaloes found at these radii are rarely satellites. As the autocorrelation function of linked subhaloes at these radii is dominated by central-central pairs, we do not apply a cut at $r \ge 2r_\mathrm{vir}$.

\bsp
\label{lastpage}
\end{document}